\documentclass{ws-procs9x6}%-tmp}

\usepackage{graphicx}
\def\gtorder{\mathrel{\raise.3ex\hbox{$>$}\mkern-14mu
             \lower0.6ex\hbox{$\sim$}}}
\def\ltorder{\mathrel{\raise.3ex\hbox{$<$}\mkern-14mu
             \lower0.6ex\hbox{$\sim$}}}

\begin{document}

\title{Hadrons in the Nuclear Medium -- \\
Quarks, Nucleons, or a Bit of Both?}

\author{John Arrington}

\address{Physics Division, Argonne National Lab, \\
Argonne, IL 60439, USA\\
E-mail: johna@anl.gov}

\maketitle

\abstracts{
Quantum Chromodynamics (QCD) is the theory governing the strong interaction
of particles. It describes the interactions that bind quarks and gluons into
protons and neutrons, and binds these into nuclei.  We believe QCD to be as
fundamental and complete as QED, the theory of electromagnetic interactions,
whose predictions have been tested to more than ten decimal places.  If it were
possible to make calculations in QCD the same way we can in QED, we would have
removed one of the biggest obstacles in the way of understanding matter in the
universe. Unfortunately, the properties of QCD make such calculations
impossible at present. Historically, there have been two approaches to
this problem.  First, we work to improve our ability to solve QCD, with the
most visible effort being the field of Lattice QCD.  Second, we make models of
QCD that attempt to incorporate what we believe to be the most important
symmetries, dynamics, or degrees of freedom, and then test these models against
experimental measurements sensitive to these assumptions.  Even the earliest
quark models of hadrons structure and the simplest bag models have had great
success, far beyond any reasonable expectation, indicating that these models
have isolated some of the key features of QCD.  More detailed models and ever
more sophisticated experimental tests are significantly improving such
details, and helping to better identify the most relevant features of QCD, one
of the key missing pieces in our understanding of the nature of matter. \\
\\
I will discuss the approach to understanding QCD by identifying the most
important missing ingredients in the simplified models we use.  I will begin by
discussing one family of simplified models: specifically, the models we use to
describe the interactions of protons and neutrons and the formation of nuclei,
and spend some time discussing the reasons that such a simplified model can be
so successful in describing nuclei.  I will then discuss a series of
experiments designed to probe some of the details of such models, aiming
specifically at regions where the simplified assumptions, specifically the
neglection of the fundamental partonic constituents of nucleons, are most
likely to break down and where new insight into the details of the full,
underlying theory of QCD will be accessible.
}

\section{Introduction to the Introductions}

As you will soon see, this is meant to be a very informal overview of
what I consider to be an extremely interesting area of nuclear physics.  The
topic I was given was ``Hadrons in the Nuclear Medium,'' but in fact it's
really part of a much broader program of describing the ``real world'' --
protons, neutrons, nuclei, and their interactions -- in terms of the ``real
theory'' -- Quantum Chromodynamics (QCD). I plan to spend a fair bit of my
time giving simple pictures of what's going on and trying to draw connections
between these different pictures. I'll then give some experimental history,
followed by a somewhat more detailed discussion of a particular set of
experiments, namely inclusive electron scattering, as way of getting at the
information we want. Keep in mind that it's really a part of a broader
program, so I will try to connect to other work that's going on.  I won't try
to be very rigorous in experimental or theoretical details, but hope to
provide at least a general overview and maybe some new intuition.  I've also
tried to provide references with nice overviews or appropriate introductions
to some of the topics that I will touch on but not discuss in detail. Finally,
many ``facts'' and numerical examples rely on Google and other equally reliable
sources, so buyer beware!

I now present two introductions, each equally true (in their own way).  If
you're like most physicists, you'll admit that they're both true, but you'll
\textit{believe} one of them.  Take your pick.

\subsection{I've got a Theory: ``Traditional'' Nuclear Physics}

As we all know, matter in the universe is made from three fundamental
particles; the proton, the neutron, and the electron.  A collection of
$Z$ protons and $N$ neutrons form bound states (nuclei) over a wide
range of $N$ and $Z$ values with $N \gtorder Z$.  These nuclei form neutral
atoms, a bound state of the nucleus and $Z$ electrons.

The proton and neutrons, collectively known as nucleons, are bound together
by the strong force of traditional nuclear physics.  This force has a very
short range, on the order of 10$^{-15}$~m, and is mediated via the exchange of
mesons.  The short range of the strong nuclear force can be understood in
terms of the uncertainty principle: the exchange particles are virtual and
have a minimum energy equal to their mass, limiting the time over which these
virtual particles can exist to roughly $\hbar / m$, and thus the maximum
distance they can travel to $\hbar c/m$.  This means that the lightest meson,
the pion ($m_\pi$=140 MeV), provides the longest range component of the
interaction with a range of order 1.5 fm.  Exchange of a heavier meson, or
multiple mesons, provide a stronger interaction at shorter ranges, including
the strongly repulsive short range core of the nucleon--nucleon interaction. Electrons
are bound to the nucleus via the electromagnetic force, due to the attraction
between particles of opposite charge.  Because the electromagnetic
interaction is mediated by the exchange of massless photons, it has an
infinite range.

Of course, some people in high energy physics or who study QCD worry about the
qu**ks and gl**ns, but they're missing the point.  A practical description of
matter in the universe requires a clear understanding of the interactions of
protons and neutrons, and how they form the nuclei that provide the core of
matter and the fuel of stars.

\subsection{I've got a Theory: Quantum Chromodynamics (QCD)}

As we all know, matter in the universe is made from three fundamental families
of particles: quarks, leptons, and bosons.  The quarks exist only in bound
states (hadrons) consisting of three quarks (baryons) or one quark and one
anti-quark (mesons).

The quarks interact by the strong force of QCD. This force is mediated by the
exchange of gluons, which couple to the color charge of the quarks (and other
gluons).  The strong force in QCD is extremely strong and, because it is
mediated by exchange of massless bosons, has infinite range.  This is similar
to QED, where the force is carried by photons that couple to electric charge,
but there are at least two critical differences. First, the coupling of
photons to electric charge is weak, with a scale set by the fine structure
constant, $\alpha \approx 1/137$, while the coupling of gluons to the color
charge is much stronger, set by the much larger strong coupling constant,
$\alpha_s \approx 0.12$ at the mass of the $Z$-boson and even larger at lower
energy scales.  Second, the gluons also carry color charge, leading to
interactions between the gluons, while the photons have no electric charge
and therefore cannot interact directly. The
combination of the stronger interaction and the self-coupling makes QCD a
non-Abelian (\textit{i.e.} very nasty) theory and yields an interaction whose
characteristics are extremely different than in QED.

The quarks are bound together into hadrons: color neutral objects that
have only residual color interactions with other hadrons. The residual
color interaction provides a weaker, shorter range force, much like the van
der Waals interaction which is a weaker version of the electromagnetic
force between two electrically neutral composite objects.

Of course, some people in nuclear physics or astrophysics worry about neutrons
and protons as something other than bound states of QCD, but they're missing
the point.  Nucleons are just convenient degrees of freedom; QCD provides
the true and fundamental description of matter in the universe.

\section{Historical Overview -- The Last 100 years}

Both versions of the introduction are correct in their own way.  From a
historical point of view, our understanding of matter was an ``unpeeling''
process.  Matter is made up of atoms, and the famous 1911 Rutherford scattering
experiment demonstrated that atoms had a small, dense, positively charged core
surrounded by a much larger and diffuse distribution of negative charge. In
1913, Rutherford discovered the proton, or more specifically, since the name
``proton'' wasn't widely adopted until several years later, the nucleus of
hydrogen: the lightest atom. By 1932, the proton, neutron, and electron
were all known, and their basic properties (mass, charge, spin) had been
measured.  This provided all of the basic ingredients for the ``modern''
picture of the atom, and for many people, the goal was to figure out the
recipe: the details of the nucleon interactions in order to have detailed
models of the formation and interactions of nuclei.

In some sense, this approach stopped the ``unpeeling'' procedure and
took the ingredients that were known in 1932 as the basic building
blocks for nuclei.  They then started the process of building models of nuclei
from the ground up.  Of course this required a much better understanding
of the structure and interactions of nucleons, including information on the
excited states of the protons. One of the most difficult components was
generating an adequately detailed and realistic model of the nucleon--nucleon
interaction, including all of the necessary spin, isospin, etc... dependence. 
The most straightforward way to obtain the interaction potentials is simple
two-body elastic scattering. However, difficulty in obtaining neutron
scattering data (in particular n--n data) makes it harder to understand the
isospin dependence, while inelastic scattering channels make it difficult to
measure the potentials at higher energies.  

One can fix the two-body force to nucleon scattering data, but this yields to
significant underbinding in nuclei heavier than deuterium. One approach is to
start with two-body interaction potentials and adjust the parameters to
reproduce the binding of light nuclei.  This yields effective two-body
potentials that can reproduce several observables, but neglects the effect of
three-body interactions. In fact, the situation is worse than this, because it
is extremely difficult to obtain three-body potentials without precise
knowledge of the true two-body interaction.  A better approach is to fix the
two-body interaction using only the N--N scattering data, and then include
three-body forces that yield the correct binding for few-body nuclei.  While
this is the most correct approach, it was not feasible early on, due to the
limitations in the N--N data, and because one needs extremely good
calculations for the few-body nuclei in order to constrain the input to the
three-body potential.

\begin{figure}[ht]
\begin{center}
\includegraphics[height=4.4in,angle=270]{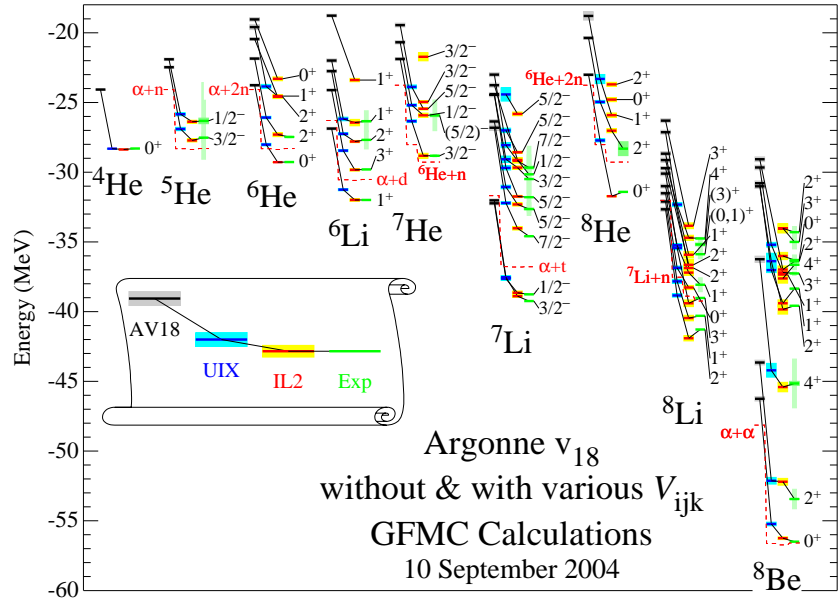}
\end{center}
\caption{Energies for ground state and excited states few-body nuclei from
\textit{ab initio} calculations based on parameterizations of two- and
three-nucleon potentials$^1$.  The left-most calculation includes only the
AV18 two-body interaction, while the next two included versions of the
three-body interaction.  The rightmost values show the measured values and
uncertainties (where available).
\label{fig:fewbody}}
\end{figure}

Over the last 20 years, we've made great progress in understanding the
interactions of protons, neutrons, and nuclei, and can now make very precise
\textit{ab initio} calculations of light nuclei, including two- and three-body
forces, and providing information on a wide range of observables: masses of
ground states and excited states, one-body nucleon momentum distributions,
two-nucleon correlations, etc.... Figure~\ref{fig:fewbody} shows the results of
few-body calculations by Pieper and Wiringa\cite{pieper01} up to A=8,
although these calculations now extend up to A=10, with some preliminary
results for Carbon.  The structure of heavier nuclei cannot yet be calculated
this way, although advanced shell model and cluster expansion approaches
exist that work well for somewhat heavier nuclei, and mean field calculations
provide some information for even very heavy nuclei. In all of these
calculations, the protons and neutrons are treated as fundamental, point-like
objects, and the structure of the proton comes in only when examine
observables where the charge or matter radius of the nucleons has to be
included. These are purely hadronic calculations, neglecting quarks, gluons,
and color, the fundamental degrees of freedom in QCD.

While this approach has been very successful, a parallel approach towards understanding
matter has also been used.  While nucleons were taken as the basic building
blocks of nuclei, it was known that they were not fundamental, point-like
particles.  The magnetic moments of the proton and neutron were not consistent
with the predictions of a point-like spin-1/2 fermion, and the differences
were at the 100 percent level.  Investigations into the sub-structure of
nucleons continued, and the idea of nucleons (and all other hadrons) being
bound states of quarks arose first as a way to categorize the observed hadrons
as excited states of three-quark systems, with the $\Delta^{++}$, composed of
three spin-aligned up quarks, showing the need for a new quantum number
(color).  The quark picture of hadrons was confirmed by deeply inelastic
scattering (DIS) measurements at SLAC\footnote{For a nice physics review and
historical perspective, I recommend the 1990 Nobel Prize
lectures\cite{taylor91,kendall91,friedman91}.}.  These measurements showed
that in high energy inclusive electron scattering, the data were consistent
with scattering from a set of quasi-free, point-like particles, moving with a
momentum distribution inside of the nucleus.  Of course they are not
free and are in fact bound tightly inside of the hadrons, but due to the
nature of their interactions they behave as if they are free, at least
compared to the scale of the energy probe being used to study them.

In what way were the data consistent with scattering from point like particles?
Well, the short answer is that the results were independent of the energy scale
of the scattering, once the energy transfer, $\nu$, and the momentum transfer,
$q$, were large compared to the other energy scales of the problem (the proton
mass and quark binding).  The idea is that in this high energy limit, the
scattering occurs on a short enough time scale that only the struck quark
is involved, and it can be treated as a quasi-free scattering in the impulse
approximation.  In this simplified case, the cross section reduces to a
convolution of the electron-quark scattering cross section and the momentum
distribution of the quark in the nucleon.  If the quarks are point-like
fermions, then we can calculate the electron-quark cross section, and if we
divide this out of the cross section, the result should depend only on the
quark momentum distribution, but be otherwise independent of the energy scale
at which one probes (usually discussed in terms of $Q^2$, the four-momentum
transfer squared).  If the quarks were not point-like, then there would be
an additional energy dependence in the cross section, and the results would
not show scaling, \textit{i.e.} not be independent of the energy of the scattering.
The scaling function or parton distribution function (PDF), basically the
cross section with the e--q cross section removed, provides the quark momentum
distribution as a function of the Bjorken variable, $x = Q^2/2M\nu$.
Scaling and the separation of energy scales are important ideas that come up
time and time again, and both will be discussed again later in these
proceedings.

After this, we quickly learned more about the partonic structure of
hadrons, and QCD, the theory governing the interactions of the quarks and
gluons.  At this point, we had all of the ingredients necessary for a
description of matter in terms of the \textit{fundamental} quark, gluon, and
color degrees of freedom.  Unfortunately, building such a description starting
from first principals is hard.  Not, ``Gosh, you'd need to be a rocket
scientist or nuclear physicist to do this'', which wouldn't have been a
problem.  No, this was much harder and to make a long story short, we still
can't make calculations of nuclear structure for even the simplest of nuclei
from first principles, and we still have a lot of work left to even describe a
single proton.

In the meantime, it's important to remember that we've learned a lot just from
trying to build such a model.  A nucleon is just a bag of three quarks, in a
constituent quark model, or three valence quarks plus a sea of virtual quarks
and gluons in QCD.  If you take two of these nucleons, it's not at all obvious
that you end up with a bound state of two nucleons, or just a single bigger
bag of quarks. This feature made for a simple way of
systematically categorizing nuclei, which was extremely important in building
up our picture of matter, but it's not obvious in advance that nature is going
to be this kind, especially in light of how nasty QCD can be.  But knowing
that nature works this way, and looking at the basic properties of QCD, we've
been able to see the importance of confinement as one of the defining
properties of QCD.

So this brings us to where we stand today.  It was straightforward (given
100 years of effort by a large number of dedicated people) to unpeel the
layers that take us from atom to nucleus to nucleons and finally quarks, but
not yet possible to take the final picture and build it back up again.  Thus,
we have two pictures of matter: the \textit{fundamental} picture of nuclei as
a complicated collection of quarks and gluons, and the more \textit{practical
and useful} picture of nuclei as a bound state of nucleons.  There
is in some sense a real inconsistency between these two pictures, and strong
opinions from people who have adopted one picture over the other as the
``right'' way to do things.  Of course, the problem is that neither picture is
at present the right way to do everything.  Understanding matter is
complicated, and the fundamental picture not always the most useful or
applicable one.  It's the reason we have physics, chemistry, materials
science, biology, etc... as courses and fields of research instead of just
physics (or Mega-Physics).  Even though it is in principle possible to use QED
to go straight from physics to chemistry, materials science, and so on, it is
certainly not the most practical way to tackle real world problems in these
fields, while in the much more complicated case of QCD, we're still working on
the simplest problems, the ``hydrogen atoms'' of hadronic physics. Even on
Star Trek, where you could tune up the engine by adjusting its quarks, gluons,
and electrons, sometimes a wrench is still the best way to reverse the
polarity\footnote{Speaking of Star Trek, and the physics thereof, I have two
things to say:  first, I have been asked more than once about funding for
fundamental research, research with no short term applications in mind, and
what kind of ``Star Trek technology'' will come out of it.  The short answer
is that I don't know what \textit\textbf{{will}} come of it, but without this
basic research, I'm willing to bet that we'll have less, and it will come
much later. Second, I'll state for the record that there's no way I'm stepping
foot in a Star Trek transporter until we know a lot more about what you and I
are made of, which is, after all, the topic of this paper.}.

So while in principle, it should be possible to calculate the structure of
matter starting from QCD, in practice this is not yet possible.  In fact, you
would lose surprisingly little by having two separate groups, working
independently of one another. The first group has never even heard of nuclei
or hadronic models of matter, and just tries to provide a quantitative
understanding of the structure/spectrum of bound states and interactions of
these bound states of quarks and gluons via Lattice, Dyson-Schwinger, etc...
approaches to understanding QCD. The second group has never heard of QCD,
quarks, or gluons, and just takes nucleons and parameterizations of their
interactions as fundamental starting points.  They can then calculate their
bound states (nuclei) and interactions starting from the ``fundamental
properties'' of nucleons (as determined by the calculations of the first
group).  The knowledge that went into generating the description of hadrons
provides almost no \textit{direct} constraints and surprisingly little insight
into how nuclei are formed from the nucleons.  For another take on this idea,
I point the reader to a recent preprint I happened across on the path from QCD
to Nuclei\cite{savage06}.

In fact, the above scenario isn't too far from the truth, in the sense that
much of the work describing nuclei was done before the knowledge of QCD or
even quarks as the constituents of hadrons. The main difference is that
instead of clever and hardworking QCD theorists, we had clever and hardworking
experimentalists who went out and \textit{measured} the properties and
interactions of nucleons -- effectively going to the ``back of the book'' to
look up the answer to the QCD bound state homework problem.  Of course, we've
had a few decades now in which people knew about both QCD and hadronic models
of nuclear physics.  But while most people in nuclear and particle physics
(and beyond) learn about both, but treat one or the other as reality when
making models or calculation, or when performing experiments using a table-top
setup, a large facility, the universe itself, or a computer as their
laboratory.

\smallskip
\noindent
$<$rant$>$ I have to stop here and say two things about Lattice QCD.  First of
all, it is an amazing tool and one day, when it reaches the point where
complete calculations using fully realistic input parameters become easy, it
will replace most if not all of what is considered nuclear physics today -
both theory and experiment.
In the meantime, while lattice QCD is providing us more and more important
glimpses into QCD all the time, I just hate it when people refer to lattice
calculations as ``experimental results.'' When I hear the word ``experiment'',
I may question the model dependence of any physical interpretation, and I
always ask questions about the experimental and analysis details, but I 
assume that the result corresponds to a world where the pion mass is very
close to 140 MeV, where there are both light and heavy quarks, quarks and
antiquarks, and while the universe may not be infinite, I at least know that I
can fit into it\footnote{I am 100\% behind the idea that it's better to have a
numerical result with some uncertainty if it means working from a model that's
closer to reality, and whose closeness to reality can be checked and improved
systematically.  But while it may be the best model around and is getting
closer and closer all the time, calling it an experiment just confuses the
issue.}.
Having statistical errors isn't enough to make something an experiment
\footnote{Please don't kill me.}.
$<$/rant$>$

\section{QCD -- ``I'm under your spell''}

Quantum Electrodynamics (QED), the theory describing the electromagnetic
interaction, has been used to make calculations good to better than ten
decimal places.  This is a remarkable accomplishment, and QED is
truly an amazing theory that essentially gives a complete understanding of
electromagnetic interactions.  In QCD, we have what we believe is the same
detailed, fundamental, complete description of the strong interaction. The
biggest difference is the fact that we just can't calculate to ten decimal
places, or even one decimal place in many cases.  In particular, the most
difficult region is where normal matter exists and interacts: low energy,
low temperature, low density.  Our inability to perform calculations in QCD
the same way we can in QED is, in my mind, the biggest single obstacle to
having a complete understanding of the form, composition, interaction, and
origin of matter in the universe. There would still be some fundamental
problems to solve that go beyond QCD: the origin of the intrinsic mass of
quarks via the Higgs mechanism, certain details about the earliest stages of
formation of matter (and antimatter) in the universe, dark matter and energy,
etc...., but these would become the final, missing pieces, which in some cases
we already think we understand, but need to verify experimentally.

One can compare the studies of QCD to some of these other problems.  For
example, we believe that the Higgs boson must exist, and that it provides the
origin of mass of the fundamental particles.  We have a theory in which we have
a lot of confidence, but we want to find the Higgs to verify the picture, and
to provide some fundamental parameters, \textit{e.g.} the Higgs mass, that cannot be
predicted in the model.  Similarly, QCD leads to the binding of quarks and
gluons into hadrons, and this binding provides 98\% of the mass of protons and
neutrons.  Again, we believe that we understand QCD as the origin of the bulk
of the mass of hadronic matter, but the theory is too complicated to solve
directly at this point.  But what we can do is take models that try to capture
the most important aspects of QCD, the symmetries, simplified models of the
binding (\textit{e.g.} confinement in bag models) and make predictions using these
models.  Experimental tests help us to evaluate these models and identify the
most relevant degrees of freedom.  This is the overall goal of Jefferson Lab,
to advance our understanding of QCD, the fundamental theory that provides the
origin of matter as we know it, by elucidating the most important aspects and
properties of this theory.

This program has several aspects, yielding different experimental programs
that often seem to be unrelated, but each of which probes a different aspect
of QCD.  The spectroscopy of mesons and baryons provides direct
input on the important degrees of freedom in hadrons. Searches for missing
hadronic states that are predicted in various quark models give input on the
relevant degrees of freedom, for example the importance of quark correlations
(di-quarks) in determining the allowed baryon states.  Similarly, the
search for hadronic states \textit{forbidden} in such quark models can
shed light on the role of glue as a fundamental constituent of hadrons.

Another aspect is detailed studies of the structure of hadrons.  This has
focussed on the proton and the neutron, as these are the most common hadrons,
and the basic building blocks of matter.  While this work is extremely
important in terms of understanding protons and neutrons as the basic
constituents of nuclei, it also provides a wealth of information on the
nature of QCD.  While we cannot solve QCD in the non-perturbative region,
nature solves it for us and the end result of QCD is contained in the quark
distributions of the proton and neutrons. So by measuring these and other
properties of hadrons, we can see the \textit{end result} of confinement by
extracting the quark momentum distributions via the structure functions and
the spatial distribution of quarks in a hadron via measurements of the form
factors. More recently, we have been able to move much closer to extracting the
flavor-dependence of both the momentum and spatial distributions, while the
new field of generalized parton distribution (GPD) studies allows more detailed probes of the simultaneous
spatial and momentum distribution.  In some ways, it is like looking at the
answers in the back of the book to help you solve a hard problem.  It can
provide insight into the problem and test solutions that you might otherwise
waste a lot of time on. While some might call this cheating, it can be a very
efficient way to test our models and our basic understanding of these terribly
complicated problems.

While I didn't specifically mention the spin structure of nucleons, and will
not discuss them any further, spin provides an additional degree of freedom
that has historically provided a great deal of information on QCD that the
more basic, unpolarized, measurements could not provide.  The same applies to
many extension of the basic measurements, form factors and parton
distributions, that have been staples of hadron structure studies over the past
decades.  Polarization, flavor-dependent measurements, measurements of the
parity violating response in electron scattering can all extend these basic
tools to provide significant new information on QCD, while the advent of GPDs
opens up the possibility to go from looking independently at the
one-dimensional spatial and momentum distributions of the quarks to a full 3-D
picture of the quark sub-structure of matter.

There are of course many other aspects of these studies that I won't address.
The pion is not a stable particle, so probing its structure is much more
difficult than probing protons and neutron.  However, the pion is one of the
simplest bound states of QCD, and it has tremendous impact on the formation of
matter.  As such, whatever we learn about the pion impacts both our
understanding of QCD and ``traditional'' nuclear physics.  Heavy mesons play
little role in ordinary matter, but they provide a system where simplified
models of QCD are more applicable, and anything we can learn about these heavy
quark systems will provide information on QCD.  While people involved in these
measurements, and many others I haven't mentioned, often focus on the direct
impact of their experiments, this is a broad, cooperative effort to understand
the physics that is responsible for the nature and form of the universe.

I've focussed here mainly on the quark sub-structure of hadrons, but QCD
goes well beyond that. If calculations of QCD were as straightforward as QED,
most of the measurements being done or planned at RHIC, Jefferson Lab, and RIA
would be unnecessary, as they could be calculated directly from QCD. There
would still be some fundamental symmetry tests, checks of the standard model,
and high-precision checks of QCD that would be important, but most of the
measurements would be unnecessary.  However, these are very different
programs, with different aims.  Much of RIA is trying to understand nuclear
structure with direct application to, for example, astrophysics and
nucleosynthesis.  This is extremely far from the physics of the underlying QCD
degrees of freedom, and so even assuming significant improvement in our
calculational tools for QCD over the next several years, we will still be
extremely far from being able to calculate the structure and interactions of
complex ($A>2$) nuclei from first principles.

The measurements at RHIC are focussed on studying the quark-gluon plasma. This
is designed to probe a specific region of QCD, based on the idea that we were
already in a position to make predictions about the nature of QCD and matter
at extremely high temperature.  These results do not provide direct
information about the structure of normal nuclei, but are aimed at providing a
new way to test our understanding of QCD in an unusual regime, where we had
tools that allowed us to better evaluate QCD. It was believed that one would
observe a weakly interacting quark-gluon plasma (QGP) in a region where it
would be easier to use QCD to describe the system.  In fact,
while there is a large body of evidence for a quark-gluon system, it is not
the naive quark-gluon plasma that had been predicted.  Instead, it is a more
complicated, strongly interacting system. While this may mean that we cannot
make some of the more quantitative tests of QCD by examining the nature of the
QGP, it clearly points to something new and unexpected that is going on at
these high temperatures, and provides us new information to help us refine our
understanding of QCD.  These measurements will provide direct information on
matter at high temperature, which may modify our picture of the very early
universe.  They may also tell us something about QCD that is relevant in other
regions, but there is much more to be done to see just what the impact of
these measurements will be.

The program at Jefferson Lab combines the practical measurements of hadron
structure, which go into our understanding of nuclei, and tests of QCD that
involve measurements designed to probe the weaknesses of our simplified models
of QCD.  There are, of course, a continuum of measurements. Any measurement of
hadronic structure provides us information against which we could test QCD
calculations, if those calculations could be adequately performed.  However,
the goal is to find observables that connect to specific properties of QCD and
which can test our models to determine the most relevant dynamics and degrees
of freedom\footnote{The web site
www.jlab.org/div\_dept/physics\_division/events/campaign\_blocks.html
provides lots of information on the physics program, providing links to 
experiments and online presentations broken down by physics topic.  A++
recommended.}.

\section{Quarks as Unimportant Degrees of Freedom}

Having but a handful of lectures to spread my philosophy and enthusiasm, and
perhaps a little knowledge, I have no choice but to focus on a very limited
part of the broader program of understanding the nature and origin of matter
in the universe. I have chosen to pick an aspect that emphasizes the role of
QCD in nuclear structure, where generally the role of QCD is usually, and quite
justifiably, ignored.  This allows us to look at certain more unusual aspects
of QCD in which we believe, but for which we have little direct evidence. In
addition, it is an area I have been involved in over the past several years,
which makes it a little easier on me, and hopefully, more useful for you.

\begin{figure}[ht]
\begin{center}
\includegraphics[width=4.4in]{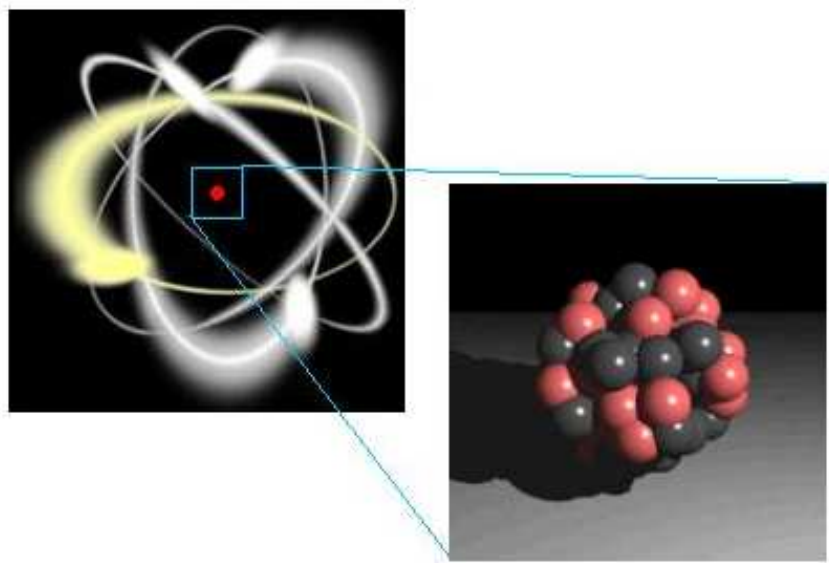}
\end{center}
\caption{The ``textbook'' picture of atom -- a small, stationary nucleus,
surrounded by a cloud of energetic electrons.  Graphics taken from
www.sciencemuseum.org.uk and www.gettysburg.edu/academics/physics.
\label{fig:atom}}
\end{figure}

Let me begin by trying to destroy some simple and common pictures that
we all learn in school.  The basic picture of matter is a nucleus, the small,
dense core, surrounded by a large, low density cloud of energetic electrons. 
In most cases, when talking about atoms we treat the nucleus as a static
and largely irrelevant object, there only as a way of localizing the electrons
which are responsible for more than 99\% of the behavior of the atom.  Now
it is certainly true that it is the electrons that matter when discussing
the interactions of atoms, be it in chemistry or material science, but this
often leads to a misunderstanding about the energy scales involved.  As I will
focus on these energy scales as a crucial component of understanding both QCD
and traditional nuclear physics, I will begin by presenting a slightly
different view of the atom, sticking to a simplified, semi-classical view for
the moment.

While the electron cloud is large, and thus is the main way in which atoms
interact with one another, the electrons provide only a tiny portion of the
mass of the atom ($<$0.1\%).  Typical binding energies (and kinetic energies)
for electrons are 10s to 100s of eV, a tiny fraction of the 511,000 eV mass of
the electron, making typical electron velocities on the order of 1\% of the
speed of light.  While this makes the electrons tremendously energetic by
macroscopic scales, they are in fact the least energetic components of atoms.
A typical binding scale for nucleons is 10s of MeV, compared to a 938 MeV
mass, making the typical velocity scale $\sim$20\% the speed of light.  Highly
energetic nucleons can have velocities \textit{\textbf{more than half the
speed of light}}, while being confined to a nucleus with a diameter of order
10 fm.  If one takes the simplest and least energetic nucleon, the deuteron,
with 2.2 MeV binding energy (yielding a velocity of $\sim$10\% the speed of light),
and imagines the two nucleons rotating in a circular orbit 4 fm in diameter, 
the nucleons would be orbiting at more than 10$^{21}$ Hz, which is pretty 
impressive.  And again, if we go deeper, we find that the quarks inside of the
nucleons are even more energetic.  The mass of the (current) quarks inside of
a nucleon is around 5--10 MeV, while the energy scale of the binding, and thus
the kinetic energies of the quarks (in this naive picture), is 100s of MeV,
making the quarks the most energetic component of matter, even though the
\textit{separation} of size and energy scales means that the quarks don't
appear to affect the structure of the nucleus, while the nucleus doesn't
affect that atomic interactions.

Now, back to the business at hand.  We \textit{know} that quarks and gluons
are the fundamental constituents of matter.  Nonetheless, models of nuclei and
low energy nuclear interactions ignore the existence of quarks entirely,
describing nuclear structure in terms of nucleons and the strong force of
``traditional'' (hadronic) nuclear physics. It is not too difficult to see why
this works, based on our qualitative understanding of QCD.  The color
attraction is incredibly strong: the force between two quarks is on the order
of 1 GeV/fm.  This is twelve orders of magnitude stronger than the Coulomb
attraction in hydrogen, and in macroscopic units this is an 18 ton force
between two quarks, each with a mass of 10$^{-26}$ grams.

Imagine a set of magnets, where two magnets are attracted by such an incredibly
strong force, and have a much weaker residual magnetic field when they are
paired up. If you're trying to separate the magnets with your bare
hands, then the kind of forces you can apply to the system won't budge the
magnets, and so it will be very effective to treat the pairs of magnets as the
fundamental objects and describe their interactions with each other and the
outside world in terms of the residual magnetic field given off by the pair. 
Of course, it's possible to describe how these pairs interact in terms of the
individual magnets, but it's going to be more complicated, and it will give
the same result in the end.  If neither you nor the interactions between these
magnet pairs can apply enough force to overcome the pairwise attractions, then
you will never have a way to separate the magnets.

Is this the correct (and complete) story for quarks in hadrons?  Well, the
interaction between quarks is incredibly strong and the residual interaction
between these color neutral bound states is much, much smaller.  The typical
energy scales in nuclei are the binding energies, which are 10s of MeV, which
are very small compared to the $\sim 1$~GeV/fm attraction between the quarks
within a nucleon.  Therefore, it would seem that the simplified analogy above
provides is a reasonable picture, and a natural explanation for the fact
that quarks do not appear as important degrees of freedom in describing
Nuclei.  You can build nuclei from protons, neutrons, and their interactions,
without worrying about the quarks inside, just as you can build a brick wall
without worrying about the fact that each brick is made up of billions upon
billions of atoms.  At least, as long as you don't accidentally smash two
of the bricks together, drop one in a furnace, or put one in a hydraulic press.

\subsection{QCD at high energy -- ``If I had a Hammer''}

Breaking the ``brick wall'' analogy doesn't take too much -- a hammer will
do the trick.  A brick will break if you hit it hard enough, in which case
you're no longer able to treat it as a ``fundamental'' building block, at
least not if you want a good wall.  Similarly, if you want to break the
picture that the nucleons are fundamental particles, you need a good
``hammer'' -- an external force (or probe) that can compete with the binding of
quarks in the nucleon, or overcome it completely.  In this case, high energy
hadron and lepton beams were, and still are, the tools of choice.

While the typical energy scales of nucleons in nuclei is 10s of MeV, one can
easily make a beam of particles with much higher energies.  This external probe
can then be used to shift or break apart the quarks within a proton, leading
to the production of excited states, the creation of additional particles, or
the complete breakup of the proton.  And these days, we have no problem making
probes that can overcome this incredibly strong binding.  An energy scale
of 1 GeV is clearly sufficient to have a significant impact on quarks bound
in a sub-Fermi object by an attractive force of 1 GeV/fm (18 tons).  So if
1 GeV is enough to overcome this force, enough to ``peek under'' the 18 ton
weight that's hiding the quarks, higher energy machines can probe deeply
into hadrons without much trouble.  For measurements at SLAC, with its 50 GeV
electron beam, we find that we're talking about pulling out a quark that's
held down by the weight of two 747s.  Higher energy facilities provide much
larger energy proton beams, where we can get at a quark that's hiding under
the Titanic (at DESY) or even ``The Iceberg'' (at LHC).  There is clearly no
problem overcoming this strong attraction, and using such high energy probes
to study the quarks inside of nucleons and nuclei. These high energy external
probes have been the dominant source of information about the quark/gluon
substructure of hadrons.

Because high energy, hard scattering is one of the regions
where we can make quantitative predictions from QCD, these experiments have
given us convincing demonstrations that QCD is, in fact, the correct theory
of strong interactions.  At high energy, the strong coupling constant becomes
weaker and QCD can be solved perturbatively.  Perturbative QCD (pQCD) makes
several predictions, including the weak energy dependence of the structure
functions extracted from deep inelastic scattering (DIS).  The DIS structure
function can be related to the distribution of quarks in the nucleons at high
energy where the scattering is dominated by scattering from a single quark
that is effectively free.  While the quarks do interact strongly, if one
has an external energy scale that is large enough, this interaction has
a negligible effect, and one can map out the quark distributions.  However,
there is still a small dependence on the energy due to the running (scale
dependence) of the strong coupling constant.  The observation of this running
provides some of the strongest evidence that QCD provides the correct
description of the color interaction, even though we can only make quantitative
predictions for QCD in limited regions.  In addition, while the running of the
coupling constant provides a test for perturbative QCD, the fact that we can
use DIS scattering to extract the quark distributions allows us to probe QCD
in the non-perturbative region.  The quark structure of hadrons comes from
quark interactions in the ``low energy'' region of QCD, where we cannot make
quantitative predictions.  By probing with these large energy scales, we can
measure the the quark distributions, thus examining the end result of the
non-perturbative physics that provides the basis for the quark/gluon structure
of hadrons.

\subsection{QCD and Hadrons at Extreme Temperature -- ``Walk Through the
Fire''}

As we've seen, the separation of energy scales leads to the impact of quarks
being well hidden in ordinary nuclei, where the energy scales are much smaller
than the quark binding, yet allows us to learn about the quark substructure by
introducing a large \textit{external} energy that can overcome this binding.
But we can also think about what would happen if the energy scales inside
of nuclei were larger.  Perhaps if the energy of the interactions between
hadrons were comparable to the interaction binding the quarks, we would find
that quarks could be easily exchanged between hadrons, yielding a breakdown
of the identity of the individual hadrons.  Perhaps hadrons would be formed
of larger systems, rather than the minimum quark-antiquark or three-quark
systems required to form a color neutral object.  These are examples of
cases where the underlying quark or gluons degrees of freedom would have an
impact on the structure of nuclei or at least large scale hadronic matter,
as the classical concept of nuclei might no longer apply.

\begin{figure}[ht]
\centerline{\epsfxsize=4.1in\epsfbox{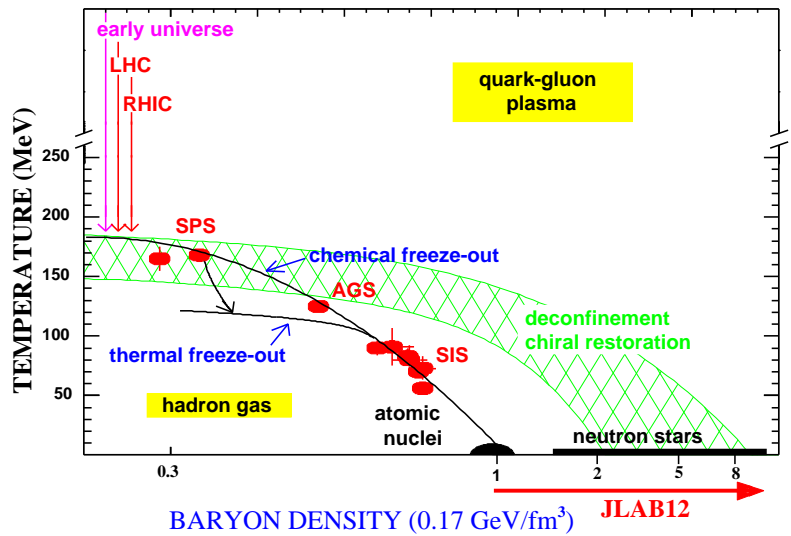}}
\caption{Phase diagram for hadronic matter.
\label{fig:phase}}
\end{figure}

\begin{figure}[ht]
\begin{center}
\includegraphics[height=3.2in,width=2.0in]{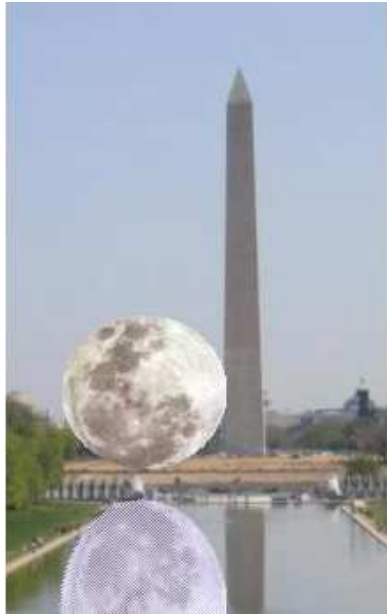}
\end{center}
\caption{The approximate size of the moon if it were as dense
as an typical nucleus.
\label{fig:moon}}
\end{figure}

While this is only a thought experiment when discussing nuclei, there are
cases where hadronic matter may have such energies, and the quark matter
of nature reveals itself.  There are two natural ways to have a large
\textit{internal} energy scale: matter at high temperature or matter at
high density.  Figure~\ref{fig:phase} shows the phase diagram for nuclear
matter, indicating the transition from hadronic matter at low temperature
and density, which we've been discussing so far, to some form of quark
matter at extreme values of temperature or density.  This transition is
expected to occur at a temperature roughly 100,000 times hotter than the
interior of the sun (300 million times hotter than the surface). 
While this is an incredibly high temperature, such temperatures existed
in the early universe, and we can try and reproduce such temperatures in
collisions at RHIC or LHC.  Densities at which this transition should occur
are several times the density of ordinary nuclei.  While this may seem
``reasonable'', at least compared to a temperature five orders of magnitude
above the sun's maximum temperature, remember that the nucleus contains
$>$99.9\% of the mass of matter, in less than one-trillionth of the volume. 
So ordinary \textit{nuclear} densities are already something like 14-15 orders
of magnitude larger than the average matter density. So matter a few times
larger than typical nuclear densities is already near or above the densities
of the densest known neutron stars (Fig.~\ref{fig:moon}).  The existence of
``quark stars'', or stars with a quark core, where the densities are
sufficient to cause the transition from hadronic matter to quark matter is
still an open and actively pursued question.

So while normally we are interested in the behavior of QCD in the low
energy region, where quarks and gluons are confined within hadrons, there
are extreme conditions under which the hadron picture may be modified and
quark degrees of freedom may come through, or in fact where the hadronic
picture becomes totally meaningless.  While we can observe neutron
stars and make models for the evolution of the universe in the first few
seconds, it's difficult to get much direct experimental information on these
systems. Of course the ideal solution would be to study matter at such extreme
conditions in the lab, where we can control the temperature and density and
where we can perform detailed measurements. For extreme conditions of density,
we're talking about densities comparable to the densest neutron stars. We
don't know of any way to produce stable matter at these densities in the lab,
and we're not likely to get funding for \textit{in-situ} studies, even on the nearest
neutron stars.  So it's not clear how we could probe such high density systems
to learn about matter at these densities.  At this point, a moderately careful
examination of Fig.~\ref{fig:phase} will reveal the notation ``JLAB12'' at low
temperature and high density, but I'll get back to this later.

On the other hand, there are clear ideas of how to probe the effect of such
extreme temperatures on hadronic matter in the lab.  The necessary temperature,
$\approx$150 MeV, corresponds to a temperature of almost two trillion degrees
Celsius (over three trillion degrees Fahrenheit for those of us who just don't
have a good intuition for Celsius).  Note that for Supernova SN1987A, there
have been estimates that the temperature in the inner layers reached 5
billion degrees, only 1000 times smaller.  I'm fairly relieved
to say that we don't know of any way to heat up a macroscopic quantity of
matter to temperatures on this scale (and I encourage readers not to try
and solve this problem, at least not in my lifetime).  However, on a
microscopic scale, say one or two nuclei, it is possible.  In a high
energy collision between two particles, some fraction of the kinetic energy
is going to be converted into heat.  In a collision between two heavy nuclei,
if the initial energy is high enough and if the interaction allows sharing
of the energy between constituents, it may be possible to generate a very
small system of extremely high temperature matter in thermal equilibrium.
This was the goal at RHIC: to collide two gold nuclei with 100 MeV per
nucleon (40 TeV total energy) and hope that for a good, solid, head-on
collision, in which a small fraction of that kinetic energy would be turned
into heat, generating a small region with temperatures that could probe the
region of this phase transition, and possibly form a quark-gluon plasma.

I won't say any more about hadronic matter at high temperature and the
quark-gluon plasma.  The reader can see the contribution to these proceedings
by Thomas Schafer, who discussed the theoretical aspects of ``The Phases of
QCD'' far better than I could, while on the experimental side, each of the
collaborations at RHIC has published a summary on their
investigations\cite{rhic1,rhic2,rhic3,rhic4}.

\subsection{QCD and Hadrons at Extreme Densities}

While high energy probes have revealed quarks and gluons as the fundamental
constituents of matter, hadronic descriptions that \textit{do not} include the
fundamental degrees of freedom of quarks, gluons, and color provide by far the
most successful models of nuclei and nuclear interactions.  The quarks
are clearly there, but the incredibly strong confinement due to the
strong color interaction in QCD conceals their presence in all but the 
most extreme circumstances.  While the presence of quarks is seen in high
energy scattering, and more recently in strongly interacting systems of quark
matter at RHIC, we do not have much direct evidence that the quark degrees
of freedom in nucleons have any impact on the structure of nuclei.

Of course, knowing that quarks and gluons are there, much effort has gone into
understanding their effects on nuclei -- trying to break down the separation of
QCD and nuclei.  One of the main approaches is to look for a change in nucleon
structure when it it placed inside of a nucleus.  We know that at extremely
high hadron densities, one goes from a hadronic state, where the quarks
interact almost exclusively with the other quarks in the same hadron, to a
quark state of matter where the identity of the hadrons is lost.  The
densities are a few or several times smaller than this in an ordinary nucleus,
but this is still a relatively large density and the presence of an
external color field -- coming from the other nucleons -- may slightly change
the binding of the quarks compared to when the nucleon is in empty space.  If
the color interaction between quarks in different nucleons has even a small
fraction of the effect as the attraction between the quarks in the nucleon, it
could yield a change in the structure of the bound state.  Perhaps the net
attraction will be slightly smaller in the presence of an external field,
yielding a slightly large nucleon, or ``nucleon swelling''.  Perhaps the
quarks nearest the neighboring nucleons might have an additional small
attraction (or repulsion) to the neighbor, yielding slightly deformed, or
maybe even ``kneaded'' nucleons.  While it might seem strange that the
nuclear environment could modify the internal structure of the nucleons,
don't forget that something similar has been observed in atomic physics. 
There is a change of almost 1\% in the lifetime of $^7$Be when the Be atom is
inside of a Buckyball due to the change in the the wave-function of the
beryllium-associated electrons\cite{ohtsuki04}.  While it makes sense that
the environment can affect the electrons enough to make a noticeable
difference, it defied the expectation that changes on the scale of the atomic
size should have a negligible effect on the nucleus, just as effects on the
energy scale of nucleon interactions in a nucleus are expected to have a
negligible effect on the quarks in a nucleon.

One can try and look for these effects in several ways, with varying
degrees of experimental and theoretical difficulties.  A conceptually
simple and clean but experimentally difficult approach is to isolate and
measure the properties of a single nucleon inside of a nucleus and look for
deviations from a free nucleon.  One can also take properties of a nucleus
as determined from a purely hadronic model and look for deviations from this
model.  Experimentally, there are many cases where one can look for such
deviations, and while any discrepancy with the hadronic description indicates
a breakdown in the model, it may be able to fix the problem by improving
the ``mundane'' nuclear physics input to the model, without resorting
to some exotic quark effect.  I'll give a quick overview of these
experimental efforts in the next section.

\section{Going Through the Motions -- A Quick History of Looking for Medium Modifications}

I'll now give a quick history of the search for medium modification. Don't let
the title get you down, as we still have a hope for a happy ending after all. More
importantly, while much of what I discuss here will be the difficulties in
interpreting data in terms of medium modification, many of these experiments
provide extremely useful information unrelated to changes in nucleon
structure, and in some cases, looking for medium modification was a secondary
(at best) goal.

I'll also point out that it is really a combined effort between experiment
and theory to try and identify modifications to nucleon structure.  In some
of the cases described below, there are significant experimental difficulties,
either direct experimental issues that were improved over time as it became
possible to make better measurements, or theoretical corrections to the
data that require good models to obtain reliable and useful data.  There are
also cases where the experimental results are completely straightforward, but
it is difficult to determine if the effect observed is best described in
terms of medium modification.  Even the issue of what does or does not qualify
as medium modification is not straightforward, as I'll discuss in the context
of some of these examples.

Finally, I want to draw a distinction between these and other searches for
quarks or quark effects in nuclei.  It is of course easy to probe the quarks
in high energy scattering, but in this case, one sees either the effect of
the quark degrees of freedom in the \textit{reaction mechanism}, when probing
the transition from hadronic to partonic descriptions of the scattering,
or else one is probing properties of the quark distributions, \textit{e.g.} the PDFs,
which are connected to the non-perturbative structure of the individual
hadrons. Both of these are important, but I am focusing here specifically on
cases where the quark effects occur \textit{because} of the nuclear medium. 
Effects generated by the internal energy scales related to the interactions of
the nucleons within the nucleus, rather than quark effects seen only because of
probing with a high energy, external probe.  So this is the effect of quarks
in nuclear \textit{structure}, rather than in the reaction mechanism.

\subsection{I'll Never Tell -- Verse 1}

Let me start by looking at experiments that have tried to isolate the effect
of the nuclear medium on an individual nucleon.  The nucleon form factors
provide information on the size and internal structure of the proton that one
might expect to change if the nucleon swells or is deformed when it resides
inside of a nucleus.  It is also a relatively straightforward quantity to
measure.  The main approach involves measuring elastic scattering from a free
proton as a function of angle for a fixed value of four-momentum transfer
squared, $Q^2$.  One can perform a Rosenbluth, or ``longitudinal-transverse'',
separation, where by looking at the angular dependence one can separate the
longitudinal and transverse components of the scattering, which are related to
the electric and magnetic form factors, $G_E^p(Q^2)$ and $G_M^p(Q^2)$, of the
proton.  The reduced cross section can be written as:
\begin{equation}
\sigma_r \equiv \frac{d\sigma}{d\Omega}
\frac {\varepsilon (1 + \tau)}{\sigma_{Mott} } =
{\tau {G_M^2}(Q^2) + \varepsilon {G_E^2}(Q^2)} \phantom{l},
\label{eq:rosenbluth}
\end{equation}
where $\tau = Q^2 / 4M_p^2$, and $\varepsilon$ is the virtual photon
polarization parameter, with $\varepsilon \to 1$(0) for $\theta \to
0$(180$^\circ$).

For a neutron, one can try and do the same thing, but because there aren't any
practical free neutron targets\footnote{At least there weren't back in the summer
of '05, during the HUGS lectures.  A new experiment, E01-015\cite{jlab_exp},
recently took data using an ``effective'' neutron target at Jefferson Lab},
measurements were performed scattering from
a neutron in deuterium, with model-dependent corrections for the binding and
Fermi motion of the nucleon, along with corrections for the proton background
in inclusive scattering, and neutron efficiency in $^2H(e,e'n)p$ measurements.
Note that the corrections for these nuclear effect in deuterium were a
limiting factor for these measurements for quite some time. This hints at the
fact that doing the same measurement for nucleons in heavier nuclei to look
for medium modification will mean being even more sensitive to these
``traditional'' nuclear effects.

The first such experiments were measurements of inclusive quasielastic (QE)
scattering, where one knocks out a single nucleon.  In this case, one is
measuring the sum of proton and neutron scattering, and so one is comparing
to the sum of the proton and neutron elastic scattering.  In addition, with
inclusive scattering, one has to choose kinematics that ensure that only
QE scattering is observed.

The simplest test is to measure QE scattering from nuclei, and compare the
total cross section for proton and neutron knockout with what one estimates
from the known e--p and e--n cross section, combined with a model that takes
into account the energy-momentum distribution, or ``spectral function'', of
the nucleons in the nucleus.  We cannot make an absolute comparison of the
cross sections, because we do not have enough information about the nucleon
distributions in nuclei.  In fact, much of our knowledge of the nucleon
momentum distribution comes from similar measurements, where we
\textit{assume} that there is no change to the e--N cross sections for
nucleons in a nucleus.  I'll discuss this approach in terms of a $y$-scaling
analysis in Sec.~\ref{sec:yscale}, and you can jump ahead and come back, or
just trust me when I say that even if we don't know the momentum distribution,
we know that if we divide out the elastic e--N cross section, what's left
should depend only on the momentum distribution.  This scaling function will
then be independent of the momentum transfer scale, $Q^2$.  This is the 
same basic idea as in $x$-scaling, where one is sensitive only to the momentum
distribution of the quarks after removing the e--q scattering cross section.
For details on the formalism, assumptions, and results I'll discuss below, the
reader is pointed to the 1990 review\cite{day90}.

If the elastic e--N cross section is changed because of a modification to the
nucleon's structure, then it can have two effects. First, it will change the
obtained scaling function, and thus the momentum distribution one would
obtain.  This isn't very helpful, because we don't have independent knowledge
of the momentum distribution for comparison (although we will see in a bit
that we can still make use of the constraint that comes from the normalization
condition for the momentum distribution).  However, it can also
introduce a $Q^2$ dependence that is not expected according to the scaling
argument given in Sec.~\ref{sec:yscale}.  This provides a test of medium
modification that does not rely on our knowledge of the momentum distribution.
This has been used to set limits on nucleon ``swelling'', because it is
sensitive to any $Q^2$-dependent change in the nucleon form factor, and the
size of the nucleon is directly related to the falloff of the form factors in
$Q^2$, at least for moderate $Q^2$ values.  The first such limit comes from
Ingo Sick's analysis of the change in the nucleon size necessary to spoil the
observed scaling\cite{sick85}.  He estimates that a change of the radius
of 7\% is about as large as one can have while still being consistent with
scaling for $^3$He, and that any such change might be twice that size for
heavier nuclei.  However, using bag models to determine the effect of changing
the size of a nucleon on other parameters, he argues that the mass should
decrease if the size increases and that the data is only compatible with a 3\%
change in size if accompanied by such a change in mass.

There is one last way to look at this data, and that is to look at the
measured spectral function itself.  We cannot compare the measured scaling
function to our expectation, because we do not know the nucleon momentum
distribution well enough to know what to expect.  However, we do know that
it should relate to the momentum distribution, and as such, the normalization
condition for the momentum distribution yields a normalization condition
for the scaling function.  By testing this, McKeown\cite{mckeown86}
was able to set an upper limit of 3.6\% for a nucleon in $^3$He, using the
same data but a different technique to set the limit.

These provide useful limits, although these limits rely on some assumptions
required for the scaling analysis, and as I will discuss in
Sec.~\ref{sec:yscale}, there are one or two particular assumptions that may
cause problems.  In addition, they rely on simple rescaling of the radius (for
both the charge and magnetization distribution), or in the case of the first
analysis, the assumption that the mass will vary as the inverse radius. In
some kinematics, one is very sensitive to the electric form factor, and thus
the charge radius, while other kinematics are sensitive entirely to the
magnetic radius.

Heavier nuclei have been measured, but not used to set limits for reasons
that I'll get to in a later section.  Well, since you insist, I'll say that
these data show a fairly clear A-dependence when examining the normalization
condition, as was done above.  The normalization condition is satisfied for
deuterium, but deviations in the normalization integral grow one goes to
carbon, iron, and finally gold, which falls 25\% short of the normalization
condition.  However, this can be explained in terms of one of those problematic
assumptions I mentioned before, and as I'll discuss later, these data appear
consistent with no swelling in an improved analysis.

\subsection{I'll Never Tell -- Verse 2}

The next level of sophistication is to perform a Rosenbluth separation to
isolate the electric and magnetic contributions. Measurements of the Coulomb
sum rule (CSR) provide this next step.  The idea is similar to the previous
measurements, except that one measures the quasielastic scattering as a
function of scattering angle at fixed three momentum, $q$, and separates the
longitudinal and transverse responses over the full QE elastic peak.
Because the tails of the QE peak overlap the inelastic region,
dominated by pion production and $\Delta$ excitations, these measurements
focus on extracting the longitudinal response, as the QE cross
section is largely longitudinal, while the inelastic backgrounds are largely
transverse.  One then takes the integral of the longitudinal strength,
the integral being less sensitive to the details of the momentum distribution,
and compares the integrated response to a calculation and looks for any
deviation that might be due to a change in the e--N cross section.  The
sum rule is basically the statement that for large $q$, the integrated
longitudinal response should just be sum of the proton and neutron elastic
scattering cross sections, and thus the ratio of the longitudinal response
to the elastic cross section should approach unity, \textit{i.e.} the sum rule
becomes saturated.

These measurements have been done for several nuclei, from $^3$He to
$^{208}$Pb.  However, they are quite challenging, as one needs a wide range of
beam energies and scattering angles to measure both small and large angle
scattering at fixed momentum transfer.  This has meant that measurements from
different facilities, MIT-BATES, SACLAY, and SLAC, have been combined to
maximize the kinematic range. One needs high precision data to
extract the longitudinal cross section from the angular dependence, which
means that differences in backgrounds and experimental details for the
different measurements have to be carefully treated and understood.
In addition, of course, is the issue of the comparison to calculation.
Using the integrated response does make the result less sensitive to some
details, but it is still important to take into account binding and nucleon
momentum distributions, especially the contribution from extremely high
momentum nucleons which we'll discuss later and which can shift strength
far enough away from the peak that it is not recovered in the experiment.

Initial results indicated significant deviations between the measurements
and calculations, indicating that the CSR was not saturated.  However,
later experiments and different analyses yielded different results,
emphasizing the issues mentioned above when combining different data sets
and looking for relative differences.  Much work has gone into this, and
the calculations seem to be fairly well under control, and most of these
experimental and analysis issues have been resolved, yielding two results:
(1) the saturation of the sum rule for all nuclei\cite{jourdan96,carlson01},
and (2) saturation of the sum rule for $^3$He, but significant (30--40\%)
violation of the sum rule for heavy nuclei\cite{morgenstern01}.  As one
can see, while most of the issues have been resolved, there are still potential
issues with the data, and more importantly, differences in the corrections
applied for Coulomb distortion -- the fact that the interaction between the
electron and the Coulomb field of the nucleus means that the electron beam
and scattered electrons are affected by the presence of the nucleus, thus
changing the apparent interaction between the electron and the struck nucleon.
This correction is in general small, but it is more important
for the relatively low energy measurements of the CSR, and the angular
dependence of the correction is very important when you need a precise
measurement of the angular dependence.

At this point, we're waiting on JLab experiment E05-110\cite{jlab_exp}, which
will cover the full range of angles in one measurement with careful checks on
the backgrounds and systematics, and on improved calculations for Coulomb
distortion.  In the meantime, CSR measurements do not yet provide a definitive
answer on medium modifications.

\subsection{I'll Never Tell -- Verse 3}

\begin{figure}[ht]
\centerline{\epsfxsize=3.4in\epsfbox{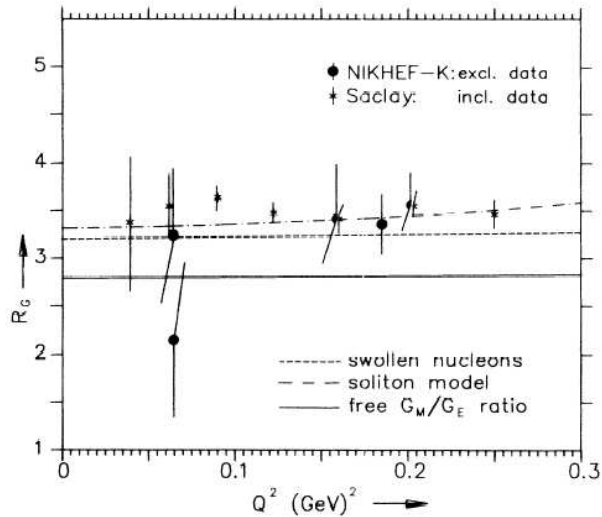}}
\caption{$R_G = 2 M_p \sqrt{(\sigma_T/\sigma_L) / Q^2}$ for proton knockout
from carbon.  For a free proton, $R_G$ would yield $G_M/G_E$.  Figure
from Ref.$^{18}$.
\label{fig:steenhoven}}
\end{figure}

A similar, but more direct, approach to performing the Rosenbluth separation
involved coincidence measurements of quasielastic scattering, where one can
clearly separate elastic from inelastic contributions, and where one can look
directly at proton form factors, rather than a sum of proton and neutron
scattering.  By measuring A(e,e'p) from the relatively light nuclei
carbon\cite{vandersteenhoven86} and lithium\cite{vandersteenhoven87},
one avoids the large Coulomb distortion corrections, while still obtaining
densities close to those for heavier nuclei.  These measurements showed,
basically, that the ratio of the transverse to longitudinal strength was
different for a proton in a nucleus than for a free proton.  From
Figure~\ref{fig:steenhoven}, we can see that the data indicated a change of
about 20\% in the ratio $G_M/G_E$.

While this had many experimental advantages, the exclusivity of the
measurement means that one must include not only the traditional nuclear
effects due to the proton binding and momentum distribution, but also
the effect of proton final state interactions.  Shortly after the
experimental result were obtained, calculations\cite{cohen87} including
final state interactions in the distorted wave impulse approximation showed
that the data could be explained in terms of final state interactions (FSI),
rather than requiring medium modifications to the proton.  Of course, this
isn't the end of the story.  Clearly, one is measuring a combination of FSI
and any modification to hadronic structure.  One needs detailed and reliable
calculations for FSI before one can conclude that one does (or does not)
observe medium modification.  So while the coincidence measurement had some
advantages, one would like to find a measurement where the final state
interactions are smaller, or else better under control.

\subsection{I'll Never Tell -- Verse 4}\label{sec:verse4}

Recently, polarization transfer measurements have been used to extract the
proton\cite{punjabi05} and neutron\cite{madey03} electric form factors.
For the neutron case, one uses the neutron in a deuteron, and has to apply
model dependent corrections for the small nuclear effects in deuterium.
For the proton one can turn this around and measure the proton form factor in
a nucleus and use the comparison to the free proton to study these nuclear
effects.  The polarization transfer technique was expected to be less sensitive 
to final state interactions and details of the nucleon distribution in the
target nucleus, and thus expected to provide a better test of in-medium
proton structure than the experiments described above.

As a side note, I should point out that the recent polarization transfer
measurements for the free proton\cite{punjabi05} disagree with the Rosenbluth
results\cite{arrington03a}.  This discrepancy led to questions about the
Rosenbluth data and even the technique, which could certainly have some
bearing on these measurements of in-medium form factors\cite{arrington04a}. 
However, at this point, there is a mounting body of evidence that this is
related to two-photon exchange
effects\cite{arrington04b,afanasev05,blunden05}.  This is good for these
measurements in two ways.  First, indications are that the effect on the
polarization transfer measurements is small.  Second, the exchange of the
second photon with the struck proton should be the same for a proton in a
nucleus, and so while the actual form factors extracted from the Rosenbluth
measurements need to be corrected for two-photon exchange corrections, the
comparison of free proton to in-medium proton should be unaffected.

\begin{figure}[ht]
\centerline{\epsfxsize=3.4in\epsfbox{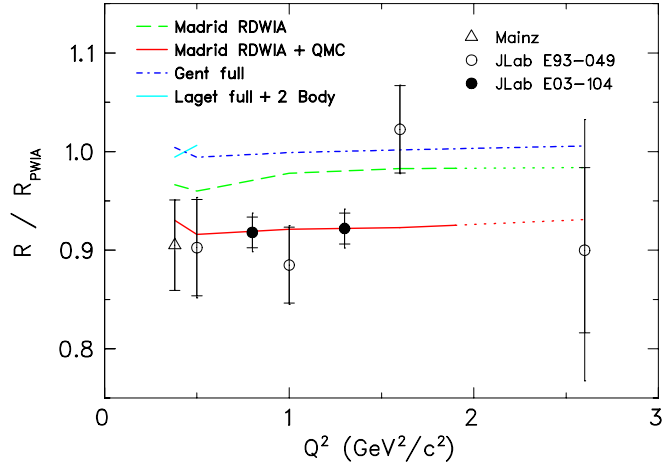}}
\caption{Comparison of the measured $R = (G_E/G_M)^{bound}/(G_E/G_M)^{free}$
with the RPWIA calculation.  The hollow points are the existing measurements,
while the solid points are projected uncertainties for an extension to the
JLab measurement.  The results are compared with several calculations of FSI,
and one calculation (solid line) including both FSI and the QMC model of
medium modification.
\label{fig:poltransmm}}
\end{figure}

Two such measurements have been completed\cite{dieterich00,strauch02},
looking at a proton in $^4$He where the peak nuclear density is relatively
large, but where the nuclear structure effects can be treated in realistic
few-body calculations.  The ratio of $R = G_E/G_M$ for a proton in $^4$He to
$R$ for a free proton was approximately 10\% below the prediction from a
relativistic PWIA calculation taking into account the nucleon binding and
motion.  All available calculations for the final state interactions indicated
that these should be 5\% or less over the entire range of $Q^2$ values
(Fig.~\ref{fig:poltransmm}).  Only the prediction of the Quark Meson Coupling
(QMC) model\cite{lu98} reproduced the results, by including the effects of
the nuclear medium on the internal structure of the nucleons.  The QMC model
has also been used to predict the effects of the nuclear medium on the quark
distributions (the EMC effect) and other observables.  It includes the
effect of the nuclear medium by allowing quarks in the nucleons to interact
via meson exchange, applied in terms of additional scalar and vector effective
fields that couple directly to the quarks.

This is a significant improvement, both in precision and interpretability,
over the similar unpolarized measurements.  However, these measurements are
still sensitive to both the normal final state interactions and the more exotic
medium modifications.  While existing calculations could not reproduce the
observed suppression of the form factor ratio, the authors were careful to
say that at this point, the calculations for final state interaction 
were not so reliable as to ensure that something beyond FSI must be needed to
explain the data.  These new data generated significant interest, as well
as new calculations which attempted to improve the modeling of FSI for this
reaction.  Recently, a new calculation\cite{schiavilla05} of the final
state interactions, including charge-exchange interactions and meson exchange
corrections, was able to reproduce the observed suppression of the form
factor ratio.  Further measurements to come from JLab (experiment
E03-104\cite{jlab_exp}) will provide better data on the suppression, an
improved measurement of the dependence of the effect on the initial nucleon
momentum, and additional polarization observables (the induced polarization).
These can be used to test this recent FSI model, to see if it can consistently
explain all aspects of the measurement, or if additional FSI effects or 
some form of medium modification may still be necessary.

\subsection{Standing in the Way}

As one can see, there are many experimental difficulties in trying to measure
the structure of a nucleon inside of a nucleus and compare it to a free
nucleon.  The other approach is to make a clean measurement of the quark
structure of a \textit{nucleus}, and then compare this to the quark
distribution we get by just taking the sum of the quark distributions of the
\textit{nucleons}. In this case, the data are clearly telling us what's
happening to the quark distributions.  However, the interpretation of this in
terms of hadronic or non-hadronic physics is what's standing in the way.

In other words, this approach takes most of the difficulties and uncertainties
and shifts them from the experimental side over to the theoretical side.  We
have to have a model of the nucleus in terms of nucleons, difficult for all
but the lightest nuclei, and combine this with our knowledge of the nucleon
quark distributions.  The proton parton distribution functions (PDFs) are well
measured but the neutron PDFs are, as usual, taken from measurements on
deuterium, where we have to correct for the nuclear effects and subtract away
the proton contributions.  Just as the previous measurements can be used to
study FSI if one neglects medium modification, or used to search for medium
modification given a model of FSIs, this data can be used to look for nuclear
effects if one knows the neutron structure, or can be used to extract neutron
structure if given a model of the nuclear effects.

To avoid some of these issues, we don't compare the nuclear PDFs to the sum of
proton and neutron, we take the ratio of a heavy nucleus, ideally with N=Z,
and compare it to the PDFs for the deuteron.  This way, we are insensitive to
the difference between the proton and neutron PDFs, and we will have at least
a partial cancellation due to the effects of binding and Fermi motion between
the deuteron and the heavy nucleus.

Searching for medium modification was not exactly the idea behind the first
such measurements.  At the time, the general expectation was that the very low
energy interactions between nucleons couldn't have any significant effect on
the quark distributions. The European Muon Collaboration (EMC) at CERN simply
wanted to compare a heavy nucleus (Fe) to deuterium to verify that a simple,
cheap, and thick iron target could be used as a effective substitute for a
thinner and more complicated deuterium target.  When they compared
muon--nuclear deep inelastic scattering from deuterium and iron, they found
that at large quark momenta, the quark distribution was suppressed in Fe,
while for small quark momenta, the distribution was enhanced\cite{aubert87}. 
This led to the reexamination of other electron scattering data from nuclear
targets, as well as additional measurements to carefully map out these
effects.  I will cover the basics of these measurements, but interested
readers should look into more details reviews of the experiments and
physics\cite{arneodo94,geesaman95}.

\begin{figure}[ht]
\begin{center}
\includegraphics[height=3.8in,angle=270]{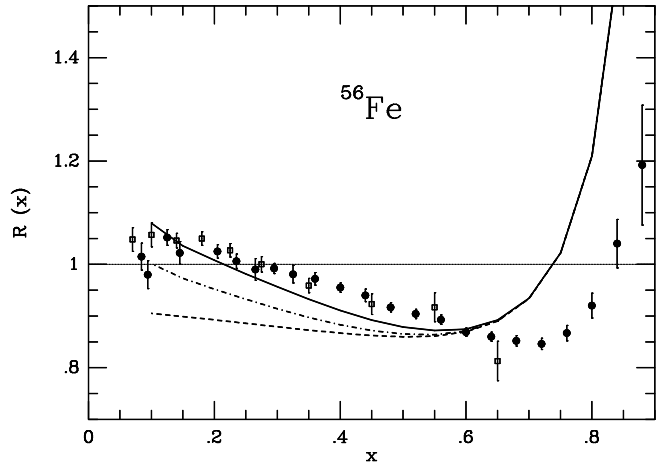}
\end{center}
\caption{Measurements of the EMC effect for Iron from SLAC E139 (circles) and
BCDMS (squares).  These are compared to a calculation by Marco, \textit{et
al.} (see text), with and with meson contributions, which takes the ratio of
iron to proton plus neutron, rather than deuteron.
\label{fig:emc}}
\end{figure}

The most complete measurement was E139 at SLAC\cite{gomez94}.  They
measured electron scattering from eight targets, from $^4$He to $^{197}$Au, for
$0.1 < x < 0.8$, and mapped out in detail the $x$ and A dependence of the
``EMC effect''.  They found that for $0.1 < x < 0.3$, there was a small
enhancement of the PDFs in the heavier targets, while for $0.3 < x <0.8$,
there was a significant suppression that was greatest around $x \approx 0.7$ (see
Fig.~\ref{fig:emc}).  The effect was larger for heavier nuclei, and scaled
approximately linearly with the average nuclear density.

Above $x=0.8$, there is a sharp rise and a large enhancement, easily understood
in terms of simple Fermi motion.  For a stationary nucleon, the PDF rapidly
approaches zero as $x \to 1$, \textit{i.e.} as one approaches the kinematic limit.  The
motion of the nucleons washes out this falloff and the amount of strength
that moves out to larger $x$ depends on the Fermi motion of the nucleus,
yielding an enhancement in all nuclei, but a larger enhancement in heavier
nuclei. Because this region is dominated by the Fermi motion, it is generally
not considered in discussion of the EMC effect.  The data at very large $x$
values has very limited statistics in this region, in part because of the 
small cross sections at the large $x$ and $Q^2$ values necessary, and in part
because of the idea that this region is not as important in understanding
what's going on.

Figure~\ref{fig:emc} shows the EMC effect measurements of $R = (2/A)
\sigma_A/\sigma_D$ for Iron, along with a relativistic mean-field calculation
by Marco, \textit{et al.}\cite{marco96}.  The bottom (dashed) curve shows
just the contribution from the nucleon quark distributions, with the
suppression at low $x$ and the increase at high $x$ due to binding and Fermi
motion.  Because the calculation is the ratio of the nuclear structure
function to the sum of proton plus neutron, and therefore neglects the effects
of Fermi motion in the deuteron, it overestimates $R(x)$ for large $x$ values.
The other curves include the quark distributions of nuclear pions (dot-dashed)
and pions plus rho mesons (solid), which I'll get back to in a bit.

First, note that looking at the data alone would indicate that something
interesting is happening for $x \approx 0.15$ and $x \approx 0.6$, while for
$x \approx 3$, there is no effect.  However, this and other calculations
indicate that much of the effect at large $x$ values is due to Fermi motion
and binding, and that this binding also has a significant role down to very low
values of $x$.  Instead of trying to explain the deviations from unity, we
really want to explain the deviations from the ``trivial'' effects of binding
and Fermi motion, which appear to be most important at lower $x$ values.  So
if we want to look for any exotic effects, we need to have a reliable model
for the effects of binding and Fermi motion to provide a reliable baseline.

This is one of the difficulties in trying to understand what we can learn from
the EMC effect.  The data at large $x$ values is often ignored, because it is
explained in terms of trivial nuclear effects.  In addition, the data for the
largest $x$ values is very limited due to the difficulty of reaching these
very large $x$ values in the DIS limits.  Therefore, little work has gone into
verifying these binding models in the region where we believe that they
provide the dominant effects, and most explanations of the EMC effect use mean
field calculations or are calculations for infinite nuclear matter, scaled down
to normal nuclear densities.  Thus, they do not include realistic input for the
nuclear structure effects.  This lack of reliable models to provide a clear
baseline for comparisons makes it unclear exactly what effect one is trying
to reproduce to bridge the gap between the binding models and the effects
observed in the data.  This was a particular problem for early explanations in
terms of more exotic effects.

The other problem is that there are different explanations that have been
suggested to explain the EMC effect, and several of them can explain at least
some of the observed effects, but there is no model that fully explains the
measure EMC ratios while being consistent with other observables.  Some
of these models explain the density dependence of the quark distributions
in purely hadronic effects, for example by including the contribution to the
PDFs coming from the quarks inside of virtual mesons in the nucleus.
Nuclear pions, the force carriers responsible for the nuclear binding, are
explicitly included and their quark (and anti-quark) distributions contribute
to the nuclear PDFs.  One example of such a calculation is shown in
Fig~\ref{fig:emc}.  However, early models that explained the EMC effect
in terms of nuclear pions also predicted a large change in the nuclear
anti-quark distributions that was not observed in Drell-Yan scattering
experiments that probe the nuclear anti-quark distributions\cite{alde90}.

Other models explain the EMC effect in terms of more exotic effects, typically
involving a modification to the structure of the nucleons in the nuclear medium.
Models of nucleon ``swelling'', multi-quark clusters, dynamical rescaling
models, etc... all proposed more exotic explanations for these measurements.
However, all of these explanations had trouble explaining all of the data, or
had apparent inconsistencies with other measurements, \textit{e.g.} the limits on
nucleon ``swelling'' discussed above.  On top of this, both the more
traditional and more exotic explanations require a good baseline, and it is
difficult to know precisely how much of the effect requires further explanation.

Even with the limitations of the data, there have been clear conclusions
made from the data.  Recently, it was shown\cite{smith02} that the
EMC effect cannot be explained entirely in terms of simple binding and
Fermi motion corrections.  It was subsequently demonstrated that the
entire effect could be explained in a binding model\cite{rozynek05}.
As you see, clear conclusions do not always agree.  Partially, this
is a question of what one means when one discusses including ``only binding'',
as there are different ways to treat the effects of binding of the nucleon
which can be more or less exotic.  In some cases, it is not even clear what
one means by an ``exotic'' explanation, or even to what extend one can draw
the distinction between a medium modification and a hadronic explanation.  A
modified hadron, for example a larger of smaller sized proton, can be
described by applying a modification to the quark distribution of the normal
proton, or it can be written as a superposition of many proton excited states.
 Now in this case, if one finds that the data can be explained by a modified
proton radius (which will have the same effect in either of the above
pictures), it might seem that it is more ``natural'' to interpret this as a
modification to the proton internal structure than as the effect of a series
of hadronic excitations that sum to the same effect, especially if it requires
an infinite sum over all possible proton-like excited states.  However, it is
not always easy to tell what explanation is more correct, when competing
pictures reproduce the same observable.  With the energy scales in a nucleus
being quite small, it seems odd to think about a modification to the quark
distributions within the proton, but it is also strange to think about
having contributions from an infinite string of excited resonances.  In the
end, people often pick what seems to them to be a more intuitive explanation.

\subsection{Rest in Peace?}

So after this somewhat bleak summary, we should decide if there is anything
else to be done, or if we should just let this question rest in peace.
It is clear that there are significant obstacles to overcome: these effects
are small, the experiments are hard, and the theory/interpretation is very
complicated.  There are things we can do to improve on these searches, such as
the polarization transfer measurements of in-medium proton form factors
described above (Sec.~\ref{sec:verse4}) which solved some experimental issues
and reduced theoretical uncertainty. Similarly, recent measurements
indicate the possibility to extend EMC effect measurements to larger $x$
values\cite{arrington03c}.  This is possible because the nuclear structure
functions show scaling at $x$ and $Q^2$ values well below the typical DIS
limit\cite{filippone92,arrington96,arrington01}, yielding the same structure
function as would be observed in DIS scattering.  This is a modification
of quark-hadron duality, which has been studied extensively for both the
proton and nuclei (see a recent review\cite{melnitchouk05} and the
contribution on duality in these proceedings for a nice overview of this
interesting topic).

In addition, several new meaurements will provide significant improvement
in the searches described above.  A new measurement of the Coulomb sum rule
(JLab E05-110\cite{jlab_exp}) will overcome some of the experimental issues
in previous measurements as well as extending these measurements to higher
$q$ values, where certain theoretical issues in the interpretation should
become less important.  For the EMC effect, new data have been taken (JLab
E03-103\cite{jlab_exp}) that will provide dramatically improved information
on the EMC effect for few-body nuclei, where several calculations show a
significant difference in the $x$ dependence, as opposed to the data for heavy
nuclei where the $x$ dependence is identical for all nuclei.  In addition, a
new round of Drell-Yan measurements (FNAL E906) will probe the nuclear
dependence of the anti-quark distributions, with better precision and a larger
kinematic range than previous data.

However, while these are important measurements, they are for the most part
providing incremental improvements over what has been done before.  They will
certainly help in our understanding, but there will still be some of the same
issues in interpreting these results.  As a rule, the experimental effects are
relatively small, and so we are sensitive to both experimental uncertainties
and model dependence in the interpretation.  Ideally, we would like to have
some new ways to probe these questions, that have the potential to provide very
clear signatures of medium modifications.

\section{Give Me Something to Sing About}

The EMC effect gives a clear indication of density-dependent effects in
the quark distributions for nuclei, although there is still some difficulty
in determining exactly what combination of effects explains this.  As mentioned
above, there have been improvements in experiments looking for medium
modification, and at the same time there have been improvements on the
calculations necessary to interpret these experiments.  For example, early
models attempting to explain the EMC effect in terms of the contribution
of virtual exchange pions tended to overpredict the effect on the anti-quark
distributions, as measured in Drell-Yan scattering.  More recent models
including meson exchange\cite{smith03} can explain the modification of the
quark distribution at small $x$ values while yielding effects on the
anti-quark distributions that are consistent with the Drell-Yan
data\cite{alde90}.  In addition, the QMC model (and other approaches) are
now able to make a consistent set of predictions for the EMC effect, nucleon
modification in nuclei, and other observables, making it easier to test
consistent evaluations of a particular model against several observables.

However, developments in experimental capabilities  and theoretical techniques 
will also open up brand new option, that promise to overcome the difficulties
that have plagued previous searches for medium modification.  For the most
part, these approaches depend on using \textit{local} density fluctuations
within nuclei to isolate regions of extremely high density.  The nucleus
is a dynamic object, and will have regions of higher and lower density.
While the \textit{average} nuclear density is a few times below the densities
where one might expect the quarks to play an important role in the nucleus, it
is possible that there may be a non-hadronic component to nuclear structure
that comes not from the average nuclear density, but from small, high density
configurations where two or more nucleons are extremely close together.

Making use of this requires several things.  We need to have a reasonable
understanding of these high density configurations, as well as a way to
cleanly isolate and probe them.  We may even be able to study the structure
of a nucleon as a function of its local environment within a nucleus, 
mapping out density-dependent modifications to its structure within a single
nucleus.  The good news is that we already know a fair bit about these
high density components to nuclei.  We have strong evidence for small
two-nucleon configurations, or short range correlations (SRCs) within nuclei,
as well as new data that studies multi-nucleon correlations.  These
configurations, where two or more nucleons are extremely close together,
provide small regions of hadronic matter at extremely high densities, and new
techniques will allow us to probe both the form factors of the nucleons and the
quark distributions of these SRCs as a function of local density, allowing us
to dramatically extend the density range over which we can look for signs of
medium modifications.  In addition, we have predictions that we may see
extremely large effects by isolating these high density systems, making
the interpretation much simpler than in previous measurements where measuring
structure for various nuclei led to a small range of densities, as well as
small experimental effects.

In the next sections, I'll discuss existing data that shows the existence of these 
high density configurations and demonstrates our ability to isolate and
study them.  I'll then discuss some experiments that will be possible with
the Jefferson Lab energy upgrade, which have the promise to provide definitive
results on density-dependent nuclear structure.  While it is missing many
recent results, there is a good review\cite{frankfurt88} that discusses these
topics, connecting SRCs to the EMC effect and medium modification, and which
was also a nice preview for much of the work that has been done in the last
15 years.

\subsection{Short Range Correlations -- High \textit{momentum} nucleons}\label{sec:yscale}

I'll start with the experimental investigations of short range correlations
(SRCs) in terms of the study of high-momentum nucleons in nuclei.  In the
nuclear shell model, or any other mean-field model of nuclei, the nucleons
are limited to momenta around Fermi momenta, $k_F \approx 250$~MeV/c for most
nuclei.  However, when one includes two-nucleon interactions, the short range
repulsive core of the N--N interaction generates nucleons at much higher
momenta.  This generates a high momentum tail that significantly changes
the momentum distribution of nucleons compared to a mean field or shell model
calculation.  Even for the lightest complex nucleus, the deuteron, the 
short range interaction yields significant contribution of high momentum
nucleons.  Close to 10\% of the nucleons have momenta above 200 MeV/c, and
these nucleons have more than 50\% of the kinetic energy.  Thus, these high
momentum components, while typically providing roughly 20\% of the momentum
distribution in heavier nuclei, are an important component of the nuclear
structure.

One important tool for studying nucleon distributions in nuclei is electron
scattering.  If one uses and electron probe and detects the scattered electron
and a single knocked out proton, the A(e,e'p) reaction, one has a kinematically
complete measurement, and can reconstruct the initial momentum and binding
energy of the struck nucleon.  To do this, one must assume that the electron
interacted only with the struck nucleon, and that the nucleon did not interact
with any other nucleons after it was struck (\textit{i.e.} absence of final state
interactions (FSIs)).  One then defines the missing energy ($E_m$) and missing
momentum ($P_m$) in terms of the difference between the total energy or
momentum of the initial state, assumed to be the initial electron and a proton
at rest, and the final state, the measured electron and proton.  This missing
energy and missing momentum then represent the initial binding energy and
momentum of the struck proton.  Alternatively, one can take the initial system
to be the electron beam and the target \textit{nucleus}, and the final system
to be the detected electron and proton, plus an (A-1) nucleus with some
recoil, taken to balance the initial momentum of the struck proton.  There are
different ways to handle the recoil (and possible breakup) of the residual
(A-1) system, but in appropriate kinematics, they yield similar results.

If one measures A(e,e'p) scattering with high resolution, one can look at the
missing energy distribution to isolate individual shells, especially in
lighter nuclei where there are fewer shells and it is easier to separate
the different shells.  One finds that these shells are not populated 
in accordance with shell model calculations; the high momentum nucleons
generated by short range interactions removes some of the strength from the
shell model distributions, yielding a suppression factor for the strength
in each shell.  By comparing the measured strength to shell model calculations, 
we can determine the ``spectroscopic factor'', basically the fractional
population of each shell relative to our expectation.  For these single
nucleon knockout reactions, the spectroscopic factors can be 0.7 or less,
indication a loss of 30\% or more of the strength relative to independent
particle shell model (IPSM) calculations.

Rather than looking for the loss of strength relative to an IPSM, one can also
use the A(e,e'p) reaction to try and directly look for nucleons with extremely
high initial energy and momentum.  However, this is made more complicated
because there are large contributions from final state interactions.
Scattering from a nucleon in one of the shells followed by a secondary
interaction between the struck nucleon and the residual nucleus yields a
background on top of the events coming from the true initial nucleon
distribution.  Even if this is a relatively small background, the momentum
distribution of the nucleons is falling off rapidly as one goes above the
Fermi momentum, and so these background processes tend to dominate the cross
section for large missing energy and missing momentum.  This makes it
impossible to use A(e,e'p) scattering to extract the energy-momentum
distribution for these high momentum nucleons without an extremely good model
for final state interactions.

When one is focussed on high momentum nucleons, one can also make use of
inclusive A(e,e') scattering, where only the scattered electron is detected. 
The advantage is that at high energy, the initial e--N scattering is decoupled
from later interactions of the (undetected) struck nucleon with the residual
nucleus.  The drawback is that one does not have enough kinematical
information to fully reconstruct the initial nucleon kinematics, and one has a
combination of e--n and e--p scattering.  However, in some kinematic regions,
in particular when looking at large nucleon momenta, one can make some
reasonable approximations and still map out the momentum distributions for
high momentum nucleons\cite{day90}.

\begin{figure}[ht]
\centerline{\epsfxsize=4.1in\epsfbox{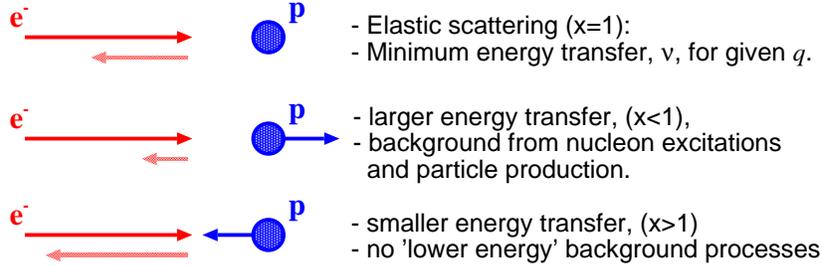}}
\caption{Electron--proton scattering at 180$^\circ$, showing the scattered
electron and $x = Q^2/2M\nu$ as a function of the initial proton momentum.
\label{fig:billiardball}}
\end{figure}

\begin{figure}[ht]
\centerline{\epsfxsize=3.7in\epsfbox{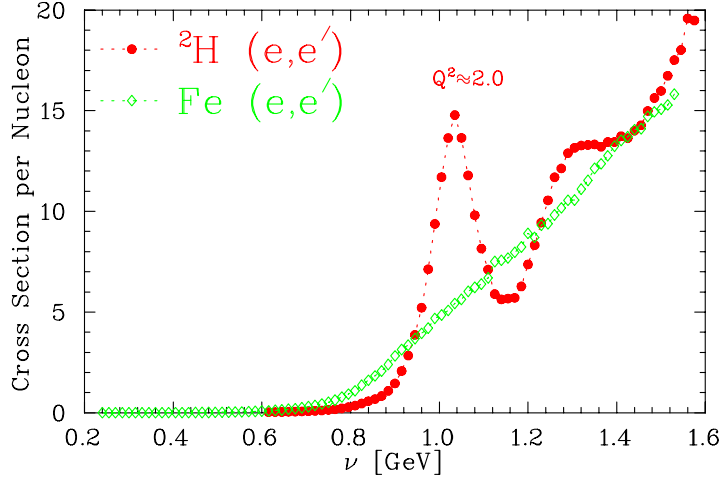}}
\caption{Inclusive cross section as a function of energy transfer for
scattering from deuterium and iron.  For deuterium, the quasielastic peak is
clearly visible, and one can see the inelastic contributions at larger energy
transfer. For iron, the QE peak is washed out, although the very low $\nu$ data
is still dominated by QE scattering.
\label{fig:qepeak}}
\end{figure}

If one starts with the limiting case of scattering at 180 degrees, as
illustrated in Fig.~\ref{fig:billiardball}, it's easy to use billiard
ball kinematics and determine the energy of a backward scattered electron
after striking a nucleon at rest.  Clearly, the electron strikes a nucleon
moving towards the incoming electron, the scattering will transfer more
momentum, and the scattered electron will have a larger backward momentum.
Similarly, striking a proton moving in the direction of the electron will
yield a lower energy scattered electron.  So in the simplified one-dimensional
case, the energy of the scattered electron provides a direct measurement of
the initial nucleon momentum.  These all have a large momentum transfer, as
the electron is scattered at 180$^\circ$, but in the case of scattering from a
proton moving towards the electron, the energy transferred is much less
because much of the momentum transfer goes into changing the direction of the
proton.  By looking at scattering with a low energy transfer, one reduces the
possibility of inelastic scattering, pion production or resonance excitation,
and can more cleanly isolate elastic scattering, as seen in
Fig.~\ref{fig:qepeak}.

The reality is, of course, somewhat more complicated, but the basic idea still
works well.  We take ($E,\vec{k}$) to be the initial electron four momentum,
and ($E',\vec{k'}$) to be the scattered electron four momentum, yielding a
virtual photon exchanged with energy $\nu = E - E'$, and momentum $\vec{q} =
\vec{k} - \vec{k'}$.  This photon is absorbed by a nucleon with
initial momentum $p_\parallel$ along the direction of the virtual photon and
momentum $p_\perp$ parallel to the photon, where the residual (A-1) nucleus
has an equal but opposite momentum. We can use energy conservation to equate
the initial energy to the energy of the final system:
\begin{equation} 
\nu + M_A = \sqrt{M_N^2 + (q + p_\parallel)^2 + p_\perp^2} +
\sqrt{M_{A-1}^{2} + (-p_\parallel)^2 + (-p_\perp)^2}.
\end{equation}
From this equation, we can see that the energy transfer, $\nu$, which
we measure by detecting the scattered electron, is related to the initial
nucleon momentum, although it includes both the parallel and perpendicular
components. Because the initial nucleon momentum is usually below the Fermi
momentum, we typically have $p_\perp, p_\parallel \ll M_N, M_A$.  Therefore,
the $p_\perp^2$ and $p_\parallel$ terms will have a much smaller effect
then the $(q+p_\parallel)^2$ term, so that we have
\begin{equation}
\nu + M_A \approx \sqrt{M_N^2 + (q + p_\parallel)^2} + M_{A-1},
\end{equation}
or,
\begin{equation}
\nu + M_N \approx \sqrt{M_N^2 + (q + p_\parallel)^2}.
\end{equation}
Again, a simple (but approximate) relationship between the energy transfer
and the initial nucleon momentum along the direction of the virtual photon.
Thus, one only has access to a one-dimensional momentum distribution, unlike
in the case of A(e,e'p) scattering, where one can measure the three-momentum.

In fact, the real world is again a bit more complicated.  First, we want to
probe high values of $p_\parallel$ to probe the high momentum tails. 
Therefore, we cannot neglect the $p_\parallel$ terms, and the equation becomes,
\begin{equation}
\nu + M_A = \sqrt{M_N^2 + (q + y)^2} + \sqrt{M_{A-1}^2 + y^2}.
\label{eq:ydef}
\end{equation}
where we have taken $y = p_\parallel$.  More importantly, we have made a
key assumption here, that the residual (A-1) system ends up in an unexcited
state.  For a low initial nucleon momentum, corresponding to a nucleon in
one of the orbitals, it is reasonable to expect that one can knock a nucleon
out of one of the shells and have an (A-1) residual nucleus that is either
in its ground state, or possibly a slightly excited state if one removed a
nucleon from one of the lower shells.  Similarly, for a deuteron target this
should be a good approximation, as one needs much more energy to excite
a nucleon excitation than to cause a nuclear excitation or nuclear breakup.

This is the typical definition used in studying ``$y$-scaling'' -- inclusive
quasielastic scattering from nucleons to probe the nucleon momentum
distributions.  In the PWIA picture, the scattering is a convolution of
the nucleon energy/momentum distribution and the elastic electron-nucleon
cross section (with complications from off-shell corrections for the
bound nucleons and final state interactions).  If one can reconstruct the
initial nucleon momentum and then divide out the electron--nucleon cross
section, one should observe scaling: the scattering should be independent
of $Q^2$, and depend only on the nucleon distribution.  The $y$-scaling
function should be identical for all $Q^2$ values, and should be directly
related to the nucleon momentum distribution:
\begin{equation}
F(y) = 2 \pi \int_{| y |}^\infty {n(p) \cdot p ~ dp }.
\label{eq:nofk}
\end{equation}
This is actually an approximate expression\cite{day90}, but it has been shown
to be a quantitatively good approximation for appropriate kinematics.
The general idea is the same as for $x$-scaling in DIS scattering; in the
region where the scattering is dominated by scattering from a single
``parton'' (nucleon or quark), the cross section is a combination of the
elastic e--parton scattering cross section and the momentum distribution of
the struck parton.

There are several limitations to this approach.  First, one does not have
information on the transverse momentum of the nucleons, but this is typically
a small effect, especially for large values of the initial nucleon momentum,
$y$, because $y^2 \equiv p_\parallel^2 \gg p_\perp^2$.  One also has to deal
with the fact that the nucleons are not only moving, they are also off-shell.
There are prescriptions for dealing with this in the cross section, the most
commonly used approaches are due to DeForest\cite{deforest83}.  However, one
also has to take the binding into account in reconstructing the nucleon
momentum from the energy conservation (Eq.~\ref{eq:ydef}).  Making the
assumption of the two-body breakup into a nucleon and an unexcited nucleus is
equivalent to using a spectral function where each value of nucleon momentum
corresponds to a single, minimal, binding energy.  To the extent that most of
the strength is near this region, this has been shown to be a quantitatively
good approximation.  As discussed above, this should be a good approximation
for low initial nucleon momenta and for deuterium, but as we will see, it is a
poor approximation for large $|y|$ values and heavy (A$>$2) nuclei.  Why is
it a bad approximation?  Because of the fact that the short range correlations
generate the bulk of the high momentum nucleons, and so the momentum of the
struck nucleon balanced mainly by a single nucleon, rather than the full
(A-1) residual nucleus.

Final state interactions also have to be taken into account, but at high
energy, the interaction occurs over a short enough time scale that the struck
nucleon does not have time to travel to and strike another nucleon.  Of
course, this will not be true if the struck nucleon is right next to another
nucleon, and we will see that this is an issue when dealing with short
range correlations.  Again, the fact that short range correlations generate
the high momentum nucleons will be an issue in using $y$-scaling to measure
the nucleon momentum distribution.

So there are three things that are required for these $y$-scaling measurements
to provide information on the nucleon momentum distribution: (1) quasielastic
scattering must dominate, (2) final state interactions must be small, and (3)
the spectator picture with an unexcited (A-1) residual state must be valid.
By making measurements at large momentum transfer (to reduce FSI), and
small energy transfer (to suppress inelastic excitations), we come close.
However, the role of SRCs in generating high momentum nucleons can lead to
a breakdown of the final two requirements.

\begin{figure}[ht]
\centerline{\includegraphics[width=3.0in,angle=0]{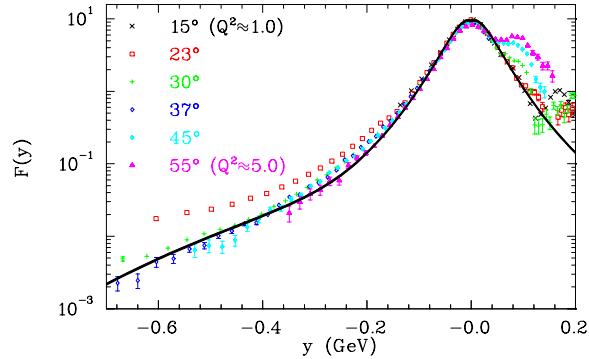}}
\caption{The extracted y-scaling function for deuteron compared to a
calculation based on the AV14 potential.
\label{fig:ystar}}
\end{figure}

To test the $y$-scaling approach, we begin by analyzing scattering from
deuterium, which provides two simplifications.  First, the reduced Fermi
smearing makes it easier to isolate the quasielastic peak as seen in
Fig.~\ref{fig:qepeak}.  Second, there is no distinction between the SRC
picture and the (A-1) spectator picture, as the (A-1) spectator is a single
nucleon in any case.  Figure~\ref{fig:ystar} shows the scaling
function\cite{arrington99,arrington01}  $F(y)$ -- essentially the cross
section with the average e--N cross section divided out -- as a function of
$y$ for data sets at several scattering angles for a fixed beam energy of 4
GeV.  On the positive $y$ side, corresponding to larger energy transfer, a
model of the inelastic contribution has been used to approximately subtract
off the inelastic scattering.  This correction is very small for negative $y$
values, as well as for positive $y$ values for the small values of $Q^2$.  The
curve is a calculation for $F(y)$ based on a calculation of the deuteron
momentum distribution using the Argonne-v14 two-nucleon potential.

From this we can see that scaling does occur for large $Q^2$ values at
negative $y$, while the lowest $Q^2$ data is above the scaling curve due
to final state interactions.  The scaling function is symmetric about $y=0$,
as it should be; without the inelastic contributions, we should be able to map
out the momentum distribution both parallel and anti-parallel to the incoming
photon.  Finally, the data at large, negative $|y|$ values is consistent with
the calculation, although the uncertainties in both the data and calculation
are at or above the 10\% level in this region, which helps to set a limit
on final state interactions in this region.  As discussed above, the
contributions from final state interactions may not vanish at large $Q^2$
for the case where the two nucleons are already overlapping.  In addition,
such FSIs may not spoil the scaling, because while the timescale for the
interaction is expected to decrease as $Q^2$ increases, it is always larger
than the timescale for the interaction of the nucleons if they are already
on top of one another.  However, while the observation of scaling does not
rule out final state interactions, the fact that the scaling function is
not dramatically increased from the expectation based on the calculated
deuteron momentum distribution in this region does set limits on FSIs in this
regime, although this data does not allow us to set precise limits.

\begin{figure}[ht]
\centerline{
\includegraphics[angle=90,width=3.0in,height=1.9in]{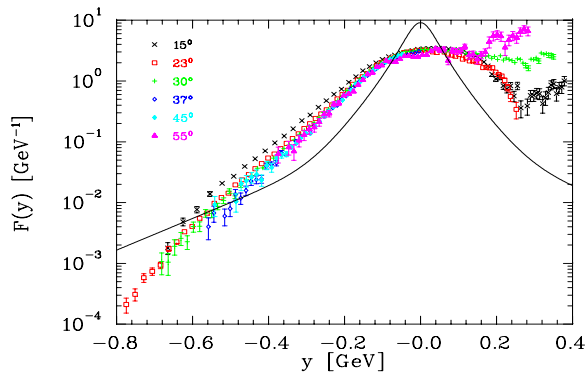}}
\caption{The extracted $y$-scaling function, compared to a fit to the
deuterium data from Fig.~\ref{fig:ystar}.
\label{fig:yfe}}
\end{figure}

However, when one goes to heavier nuclei, even $^3$He, one observes several
problems with the results, in particular for heavy nuclei and large, negative $y$
values.  Figure~\ref{fig:yfe} shows the $y$-scaling function for iron from
the same experiment, after subtracting the inelastic contribution, compared to
a fit to the $y$-scaling function for deuterium. First, we can see from the
falloff of the scaling function at large $|y|$ is much \textit{faster} in
heavy nuclei than in deuterium.  This is not consistent with the fact that the
heavier nuclei have a larger Fermi motion, nor with the picture of two-nucleon
interactions -- which should occur in heavy nuclei as well as deuterium --
generate the large momentum distributions.  In addition, the peak is clearly
asymmetric about $y=0$, although the uncertainty in the subtraction of the
inelastic background is a significant concern for large $Q^2$ values.  Finally,
as mentioned in a previous section, the scaling function extracted should be
consistent with the fact that it is related to a momentum distribution which
has a known normalization\cite{mckeown86}.  While this is satisfied for
deuterium, the scaling function falls short of the normalization condition by
more than 20\% for heavy nuclei.  One possibility is that final state
interactions are larger than expected.  However, as mentioned above, the high
energy scattering should occur on a short enough time scale that the struck
nucleon cannot interact with the other nucleons, unless they are essentially
right on top of each other.  So this should only occur when scattering from
one of the SRCs. However, if these effects are large, they should yield a
large deviations from our models of the momentum distribution in deuterium,
and the existing data are consistent with calculations, although with limited
precision.  In addition, this should only increase the strength of scattering
for large $|y|$, and yet we find a suppression of strength in heavy nuclei,
relative to what we expect.

\begin{figure}[ht]
\centerline{
\includegraphics[width=3.0in,angle=0]{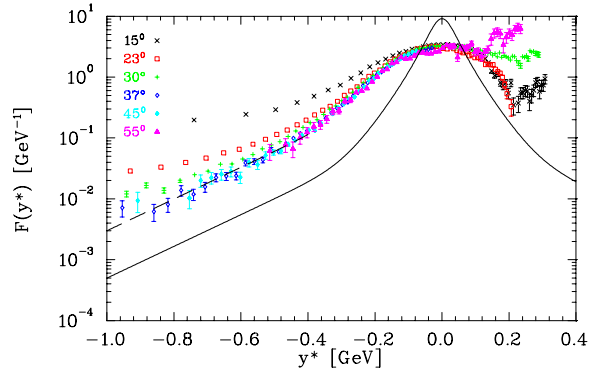}}
\caption{The extracted $y$-scaling function, compared to a fit to the
deuterium data from Fig.~\ref{fig:ystar} (solid), and a scaled up version
of the tail of the deuterium fit (dashed).
\label{fig:ystarfe}}
\end{figure}

This points to the the importance of SRCs at large momenta as the likely
suspect.  We do not include SRCs in our picture for the usual y-scaling
analysis.  By including this in our y-scaling analysis, we can try to verify
our picture of SRCs yielding the high momentum nucleons and, hopefully,
fix the problems we observe. One option is to take the dominance of
two-nucleon SRCs in reconstructing the initial nucleon momentum.  The simplest
option would be to modify Eq.~\ref{eq:ydef} using the approximation that it is
in fact a single nucleon balancing the momentum of the struck nucleon, with
the residual (A-2) system at rest and in an unexcited state:
\begin{equation}
\nu + M_A = \sqrt{M_N^2 + (q + y)^2} + \sqrt{M_N^2 + y^2} + M_{A-2}.
\label{eq:ydef2}
\end{equation}
For large values of $|y|$ and heavy nuclei, the difference between the
mean field picture and the SRC picture (Eqs.~\ref{eq:ydef} and~\ref{eq:ydef2})
can be large, significantly changing the scaling function and thus the
momentum distribution one would extract.  This neglects the motion of
the SRC in the mean field of the nucleus, which can be an important effect.
This has been addressed in two ways, by applying a mean excitation energy to
the (A-1) system that takes into account both the 2N SRC and the center of
mass motion of the SRC\cite{ciofi99}, or by explicitly including an average
momentum for the (A-2) system\cite{arrington03ystar} in Eq.~\ref{eq:ydef2}.
Figure~\ref{fig:ystarfe} shows the scaling function using the alternate
definition of $y$ taking the latter approach, along with the fit to the
deuterium data (solid line), and the tail of the deuterium distribution
scaled up by a factor of six (dashed line).  One can see that now the peak is
symmetric about $y=0$, and that the high-momentum tail has the same falloff
for iron and deuterium, although the magnitude is larger for iron.  In
addition, the scaling function is now consistent with the constraint coming
from the normalization condition of the nucleon momentum distribution.

It turns out that treatments of this kind resolve both the too-rapid falloff
of the momentum distribution for heavy nuclei\cite{arrington03ystar}, and
the problem with the normalization.  With this, we have a fully consistent
picture of what is contributing to the nucleon momentum distributions, and
a way to extract the distribution of high momentum nucleons in nuclei.  Of
course, there is still some model dependence due to the fact that we neglected
final state interactions and the fact that we use a modified version of the
three-body breakup picture of Eq.~\ref{eq:ydef2}, which provides a model for the
mean excitation energy as a function of nucleon momentum, rather than
incorporating a full two-dimensional energy-momentum distribution.  However,
to the extent that one can test these assumptions by testing the consistency
of the extracted momentum distributions (shape and normalization), the results
are consistent with these effects being small. A new experiment,
E02-019\cite{jlab_exp} was completed last year, and should significantly
extend these studies of high momentum nucleons and SRCs, with a focus on
light nuclei, $^2$H, $^3$He, and $^4$He, where we can perform much better
tests of these assumptions by comparing the data to detailed few-body
calculations.

However, for the next step, the main conclusions we need to draw from this
data is that we can isolate scattering from 2N SRCs (or even multi-nucleon
SRCs) in inclusive scattering at $x>1$.  This will allow us to isolate these
high-density components of nuclear structure and perform additional studies
to look for the important of quark degrees of freedom at high densities. We
can use the existing data to make additional tests of the dominance of SRCs at
large $x$, without relying on the assumptions of the scaling analysis
presented above.

\begin{figure}[ht]
\centerline{\epsfxsize=3.7in\epsfbox{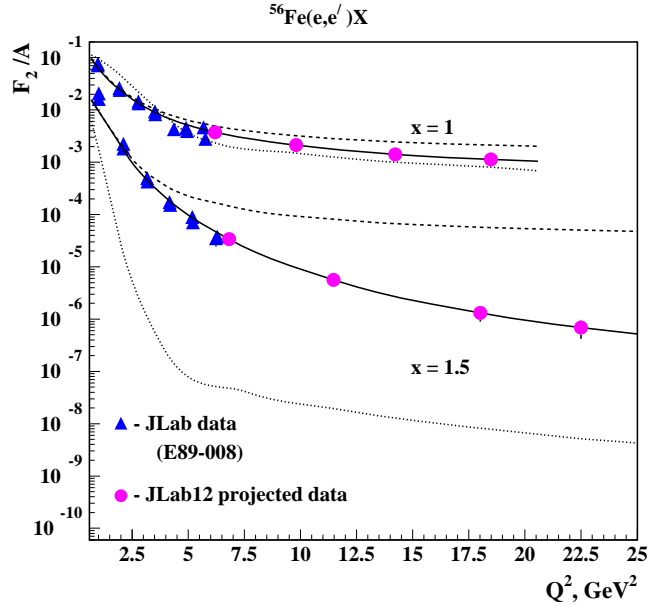}}
\caption{Cross section for Fe(e,e') at $x=1$ and $x=1.5$, compared to
existing data (blue triangles) and projected
results for measurements with the JLab energy upgrade.  The dotted lines are
mean field calculations, while the solid lines include 2N SRCs and the dashed
lines included three- and four-nucleon SRCs.
\label{fig:sargsian}}
\end{figure}

First, we can directly compare the cross sections to detailed calculations of
inclusive scattering with and without SRCs included in the model of the nuclear
structure, to see the importance of the correlations. 
Figure~\ref{fig:sargsian}, shows the cross section for inclusive scattering
from Iron at $x=1$ and $x=1.5$\cite{arrington99}, with calculations
from\cite{sargsian03}. It is clear that for large $x$, one is dominated by
scattering from the two-nucleon SRCs.  There is also little indication for
multi-nucleon SRCs in this data, at least at the level predicted by this
particular model of 3N and 4N SRCs\cite{frankfurt80}.  However, as the figure
shows, the sensitivity becomes much larger at higher energy.  As with the
y-scaling analysis, this relies on a comparison of the data to a calculation
of the cross section, but this direct comparison allows a much more realistic
calculation than the PWIA model (and assumptions about the kinematics of the
breakup) that go into the scaling analysis.

\begin{figure}[ht]
\centerline{\epsfxsize=3.7in\epsfbox{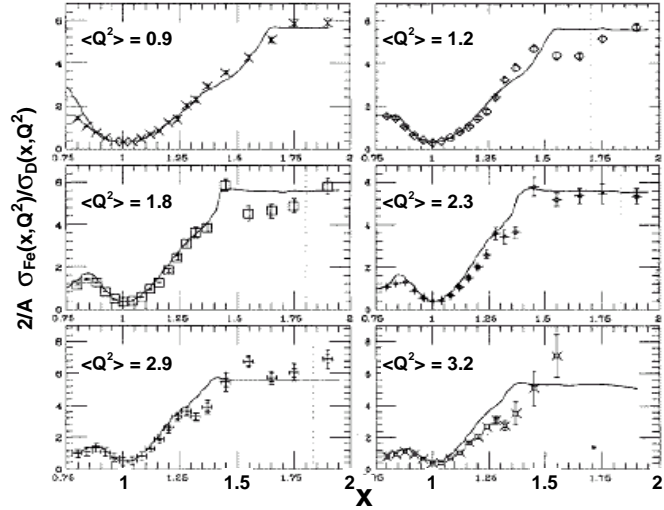}}
\caption{Ratio of iron to deuteron cross section per nucleon vs. $x$
for a range in $Q^2$ from SLAC measurements.  Figure from Ref.$^{52}$.
\label{fig:ratio}}
\end{figure}

\begin{figure}[ht]
\centerline{\epsfxsize=3.1in\epsfbox{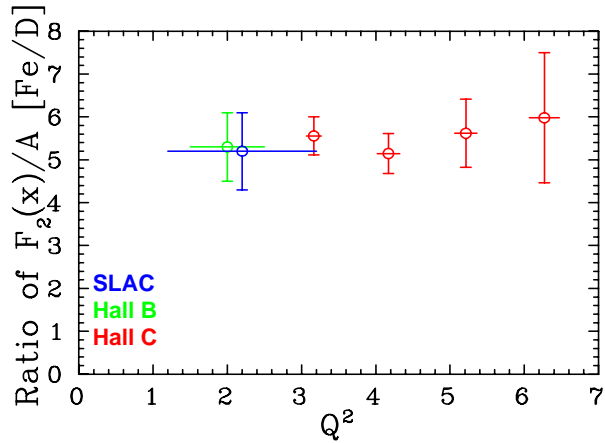}}
\caption{Ratio of large-$x$ iron to deuteron cross section per nucleon as
a function of $Q^2$.
\label{fig:ratiovqsq}}
\end{figure}

One can also look for SRCs in a more model-independent way, by taking
the ratio of the cross section (or structure function) of heavy nuclei to
deuterium in the region where SRCs are expected to dominate.  This was first
done by combining measurements on different targets from SLAC
experiments\cite{frankfurt93}.  Later dedicated measurements took
data for both deuterium and heavy targets\cite{arrington99} or $^3$He and
heavy targets\cite{egiyan03}.  All found the same results, as predicted by
the SRC picture\cite{frankfurt88}: When one is at large enough $x$ and $Q^2$,
where scattering from the mean-field nucleons becomes negligible, all nuclei
yield the same $x$ and $Q^2$ dependence, with only the normalization (related
to the number of nucleon pairs) varying from nuclei to nuclei.  One examines
this by taking the ratio of heavy nuclei to deuterium, as shown in
Fig.~\ref{fig:ratio} for the SLAC data, and observing that the ratio is
constant above $x\approx 1.5$.  Figure~\ref{fig:ratiovqsq} shows the ratio
integrated over the plateau region, from $x \approx 1.5$ to $x=2$, as
extracted from the SLAC\cite{frankfurt93} and Jefferson Lab Hall
B\cite{egiyan03} measurements, and the ratio as extracted from the Hall C
data\cite{arrington99,arrington01}.

This is a natural enough result, if 2N SRCs dominate the high-momentum
components in all nuclei. While final state interactions may be relevant, they
should be nearly identical in different nuclei as only the FSI between the
correlated nucleons should contribute, and so should largely cancel in the
ratio.  The large mean field motion of the pair in heavier nuclei will modify
the momentum distributions for large nucleon momenta, but as the falloff is
roughly exponential, a small broadening of this distribution will still be
exponential and yield only a rescaling of the ratios, rather than a difference
in the shape for heavier nuclei.  One can also extend such measurements to
look for 3N SRCs, by going to $x>2$ and comparing the ratios of heavy nuclei to
$^3$He.  First results for this are coming out of Hall B at Jefferson
Lab\cite{egiyan05}, and these measurements will be extended to higher $Q^2$
and better statistics with the Hall C data from E02-019\cite{jlab_exp}.

A recent review article\cite{sargsian03} discusses in detail additional
ways to study two-nucleon and multi-nucleon correlations.  In addition to
detailing the results discussed here and future possibilities with inclusive
scattering, it also discusses the importance of high-$Q^2$ coincidence
measurements, as well as the possibility of using ``tagged'' measurements
to more directly probe two-nucleon correlations by performing coincidence
$A(e,e'p)$ measurements with the additional detection of the high-momentum
spectator nucleon from the SRC.  New results for both of these have come
from recent experiments at Jefferson Lab\cite{rohe04,niyazov03}, and
a new measurement was recently completed (E01-015\cite{jlab_exp}) using
a ``tagged'' neutron target to study the ``Barely off-shell'' neutron.
In addition, there are significant new possibilities and distinct advantages
to the experiments that will become possible with the JLab energy upgrade,
in particular for these tagged neutron measurements.

\subsection{Short Range Correlations -- High \textit{density} configurations}

From the studies described above, we know that the high momentum components 
in nuclei are generated by two (or more) body correlations.  In this section,
we will talk about using the high momentum nucleons to tag these correlations,
which also represent local regions of high density matter.  While
\textit{average} nuclear densities are a a few times smaller than the
densities at which we expect to see a phase transition to quark matter, these
\textit{local} high-density fluctuations may provide densities high enough
that the underlying quark structure of matter is modified.  As these high
density configurations are local in both time and space, one does not expect
to have an equilibrium system or to observe a true phase transition. However,
the quark effects can be viewed as a first indication of how the hadronic
structure of matter may break down.  Given that we don't have any way to form
an equilibrium system at densities sufficient to probe the expected phase
transition, this may be our best option for supplementing the limited
information we can gain on dense hadronic matter from observation of compact
astronomical objects such as neutron stars. Ideally, any effects can be
observed cleanly enough that microscopic descriptions of these effects can be
tested and evaluated.  Plus, it would be kind of cool to be show that a small
part of the person sitting next to you (or the paper that you're reading) is
denser than a neutron star.

The first question is whether or not we expect the densities in the SRCs to be
high enough that we might see modification to hadron structure due to the
dense nuclear medium. First, we have some evidence from the EMC effect that
there is a density dependence to the quark distributions in nuclei, which may
go beyond the simple effects of nucleon binding.  One possibility is that this
is related to nucleon modification within these small, high-density
configurations, rather than an effect of the average density. If so, then one
might expect a very different behavior in few-body nuclei, compared to heavier
nuclei where the effect is saturated.  This idea will be examined with the
data recently taken by E03-103\cite{jlab_exp}, a measurement of the EMC
effect for $^3$He and $^4$He.  It can also be tested by looking at the EMC
effect in deuterium by comparing the nucleon structure observed for
low-momentum nucleons, which are nearly on-shell, and for high-momentum
nucleons which are part of a smaller sized, higher density configuration of
the deuteron\cite{sargsian03}.  Performing such measurements in the region
where the EMC effect is large will require measurements at 11 GeV at Jefferson
lab, but already measurements have been made at lower energies, comparing
scattering from high-momentum nucleons, as tagged by a backward going high
momentum spectator in CLAS\cite{klimenko05}, and scattering from low-momentum
nucleons, using a new recoil detector to tag backward going, low-momentum
spectators in E03-012\cite{jlab_exp}.  This, by the way, is currently being
used as the effective free neutron target that didn't exist back in June
during the HUGS meeting.

\begin{figure}[ht]
\centerline{\epsfxsize=3.0in\epsfysize=2.5in\epsfbox{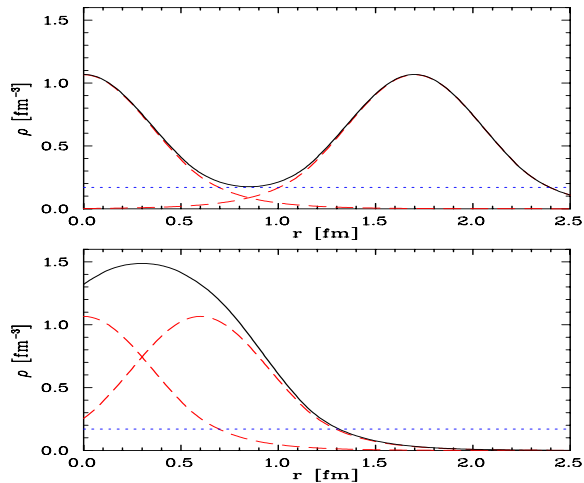}}
\caption{One-dimensional density profile for two nucleons at 1.7~fm separation,
typical in heavy nuclei, and for 0.6~fm.  The dashed lines are the density
profiles for the two nucleons, and the solid line is the sum.  The dotted
line indicates the average nuclear density.
\label{fig:twonucleons}}
\end{figure}

We can also see from simple estimates of the densities in SRCs that peak
densities may well exceed the QCD phase transition at high density.  Taking
the spatial charge distribution for a proton, as determined from the
electric form factors\cite{kelly02}, we can plot the density for two nucleons
as a function of separation, as shown in Fig.~\ref{fig:twonucleons}.  For
a typical separation of 1.7~fm, the overlap region is small, and densities
in the overlap region are comparable to average nuclear densities.  However,
for a separation of 0.6~fm, there is a large overlap region with average
densities many times ordinary nuclear matter -- as large or larger than neutron
stars, and certainly probing the region of expected phase transition from
Fig.~\ref{fig:phase}.  The extremely sharp rise of the short range repulsive
core of the nucleon--nucleon interaction occurs at around 0.4~fm, and so while
0.6~fm would be a very small separation, it is not unreasonable.

The next question is how to go about probing the internal structure of one of
the nucleons in one of these high density configurations?  First of all, since
we're mainly looking at two-nucleon correlations, there's no particular need
to go to heavier nuclei, as the 2N SRCs will be similar in all cases.  So we
can limit ourselves to the deuteron, at least for the moment.  One option that
was mentioned above is to strike one nucleon, and then detect the other
nucleon.  If it was simply a spectator, than its measured momentum provides a
measure of its initial momentum, and thus the initial momentum of the struck
nucleon.  So we basically have scattering from a single nucleon, whose initial
momentum is known, and we can measure its properties (\textit{e.g.} form factors) as a
function of it's initial momentum, and thus compare a basically free nucleon
to a nucleon in a high density configuration.

The problem is that if we strike one nucleon, the probability that it was
in a high momentum, high density configuration is small because the momentum
distribution falls off very rapidly.  Thus, there is only a small probability
that we will find the spectator nucleon at very large momentum.  In fact, if
we do find that the second nucleon has a large momentum, there is a much
greater chance that it is not a spectator.  More likely, the nucleon that
was struck then scattered from the second nucleon, yielding a large momentum
nucleon that is not truly a spectator.  Fortunately, such rescattering will
usually yield a second nucleon that is going forward, and almost never a
backward going nucleon.  So by looking for large momentum nucleons in the
backward region, one has the best chance to isolate scattering from a
high density configuration\cite{sargsian03,klimenko05}.  While there
is certainly theoretical uncertainty associated with such measurements,
due to things such as background contributions and off-shell effects, one
is looking for much larger experimental signatures of nucleon modification,
and one can look at both the $x$ dependence and the nucleon-momentum
dependence of the effects to evaluate models that attempt to describe
both these results and the existing EMC effect data.

\begin{figure}[ht]
\centerline{\epsfxsize=3.0in\epsfbox{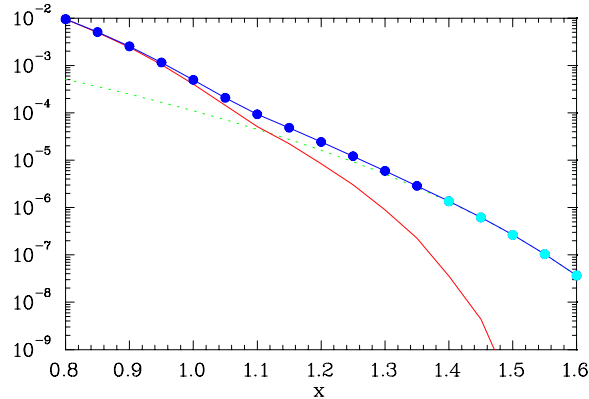}}
\caption{Parton distribution for superfast quarks in a deuteron.  The bottom
solid line is the distribution one obtains from a convolution of the
nucleon and quark momentum distributions.  The top line is what one obtains
by taking a 95\% contribution from the convolution picture, combined with a
5\% contribution from a six-quark bag model (shown as the dotted line).
The circles are projected measurements at 11 GeV, and the dark circles
indicated the kinematics that are expected to be well in the inelastic
region, and thus directly interpretable as the quark distributions.
\label{fig:superfast}}
\end{figure}

However, there is a simpler way that can still provide useful information 
about the structure of these high density configurations.  We demonstrated
in the last section that inclusive scattering at large $x$ is dominated
by scattering from the high momentum nucleons in SRCs.  At the energies
of the previous measurements, this was mostly quasielastic scattering from
these high momentum nucleons, and the measurements are sensitive to the
high momentum tails of the nucleon momentum distribution.  However, with
higher energy beams, one can still probe the large-$x$ region, but with
scattering dominated by inelastic processes.  At high energy energy, the
inclusive process becomes scattering from quarks, and one can extract the PDFs
for the high momentum tails of the quark distributions.  In the standard
hadronic picture of the nucleus, this would just be a convolution of the
momentum distribution of the nucleon in the nucleus with the distribution of
the quarks in the nucleon.  The high $x$ distribution would simply be
measuring the highest momentum quarks in the highest momentum nucleons, which
would yield a very rapidly falling distribution quark distribution for these
``superfast'' quarks above $x \approx 1.2$ (red line in Fig.~\ref{fig:superfast}).

However, if the two nucleons are close enough together that the quarks
inside can interact, this provides a new way to generate high momentum quarks
which will dramatically enhance the distribution of superfast quarks.
Any kind of quark interaction of quark exchange between the nucleons could
provide such an enhancement, and by extracting these quark distributions,
we can look for this kind of enhancement.  Figure~\ref{fig:superfast} compares
the distribution for the convolution calculation (bottom solid line) to the
effect if one has a 5\% component coming from a six-quark bag configuration
(top line).  While there is some uncertainty in the convolution model, 
based on our knowledge of the high momentum tails of both the nucleon and
quark distributions, this simple estimate predicts a huge enhancement over
the convolution model for $x > 1.2$.  While the six-quark bag is just
one simple model for describing the possible structure for the case where
one has two highly overlapping nucleons, the excess of superfast quarks is
a general feature that comes from the direct interaction between quarks in
the nucleons beyond what is included in the hadronic picture of the N--N
interaction.  If we observed this kind of dramatic enhancement of the
distribution of superfast quarks, it will provide a clear signature for a
breakdown of the purely hadronic picture of nuclei.

\section{Where do we go from here?}

Recent new measurements and future possibilities with higher energies at
Jefferson Lab will not only provide improvements on traditional methods
of looking for quark degrees of freedom in nuclei, they will also provide
completely new ways to medium modification. This information can provide new
insight on QCD, as well as telling us more about nuclear structure at extremes
of energy and density.

However, this is just one aspect of probing the deeper connections between
QCD and matter in the universe.  I have focussed exclusively on looking for
the effects of the quark degrees of freedom that are the basis of QCD but
are not part of our traditional hadronic pictures of matter.  However, 
there are similar ideas for new ways to look for the direct impact of
gluons or color degrees of freedom on nuclear structure.  There have been
many studies of color transparency\cite{jain96,garrow02,dutta03b}, the
reduction of the strong interaction due to cancellation between the color
charges when the hadron is in a small sized configuration.  These are meant to
study an exotic property of hadrons that is directly predicted by and
connected to the color degrees of freedom in QCD.  Even more so than with the
program described here, JLab has already significantly advanced such
measurements, and will provide critical extensions to these studies with the
energy upgrade.

Finally, the fundamental question of how the structure and interactions of
hadrons come about from the quark and gluon basis of QCD is another key
component, and this is also a large part of the current and future JLab
program.  Searches for missing resonances will help us understand which states
predicted by constituent quark models do not occur in the real world, thus
providing important information on the missing dynamics or symmetries in these
models.  Similarly, looking for exotic states that do not occur in these
models, be it quark states with forbidden quantum numbers or glueballs, can
also give extremely important information on key aspects of QCD in the
confinement region.  Understanding the QCD origin of nucleons and nuclei is 
one of the most central goals in nuclear physics, and a carefully laid plan
including some of the most critical measurements can provide an important road
map to those who are attempting to provide one of the most important missing
pieces in our understanding of matter: the ability to calculate or describe
QCD in the non-perturbative regime.

\section{Once more, with feeling}

While we believe that QCD provides the fundamental and complete description
of the strong interaction, the biggest limitation in our understanding
of matter is our inability to use QCD to predict the structure of hadrons
in the non-perturbative region.  Much of the work in nuclear physics, both
experimental and theoretical, is connected to trying to better understand
and evaluate QCD, either to improve our understanding or ability to calculate
in the non-perturbative region, or to test predictions in regions where
we believe that we have a more complete understanding of the consequences
of QCD.  This includes regions where the interaction is weaker, high
energy or extreme temperatures or densities, or cases where we can make
a perturbative expansion of the QCD Lagrangian in terms of other small
parameters, for example in chiral perturbation theory, where quark masses and
hadron momenta are small parameters compared to the hadronic scale.  While
the experiments seem to cover a wide range of topics and superficially have
little direct connection to one another, they have a common theme of
improving our understanding of QCD.  

One important aspect of this broad program is focused on trying understand
hadronic structure and hadronic interactions in terms of their fundamental
QCD origin. I have tried to give a broad overview of this program, discussing
connections between the QCD basis of matter and the highly effective picture
we have in terms of protons and neutrons.  My aim was to avoid losing the
forest for the trees, and I have to hope I haven't lost too many people in the
forest while doing so.  On the other hand, part of the aim was to present some
topics in a new way, and hope that it brings out new connections that weren't
obvious before.  There's something to be said for spending a little time lost
in the forest and thinking about the universe every once in a while.

\section*{Acknowledgments}

\begin{figure}[ht]
\centerline{
\epsfxsize=1.98in\epsfbox{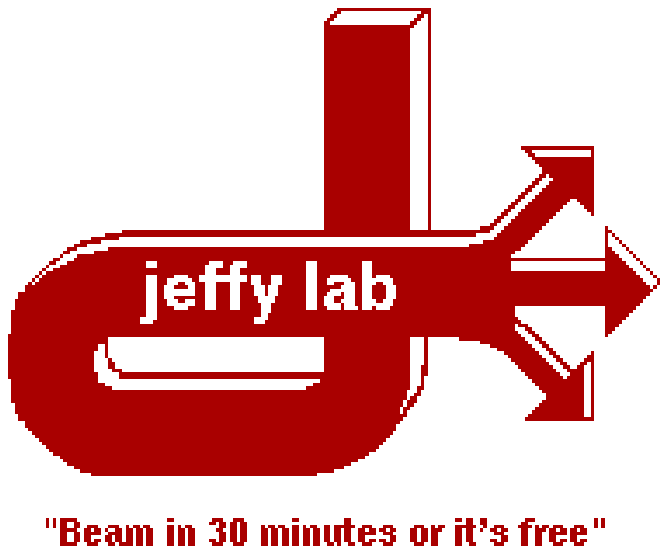} \\
\epsfxsize=2.3in\epsfbox{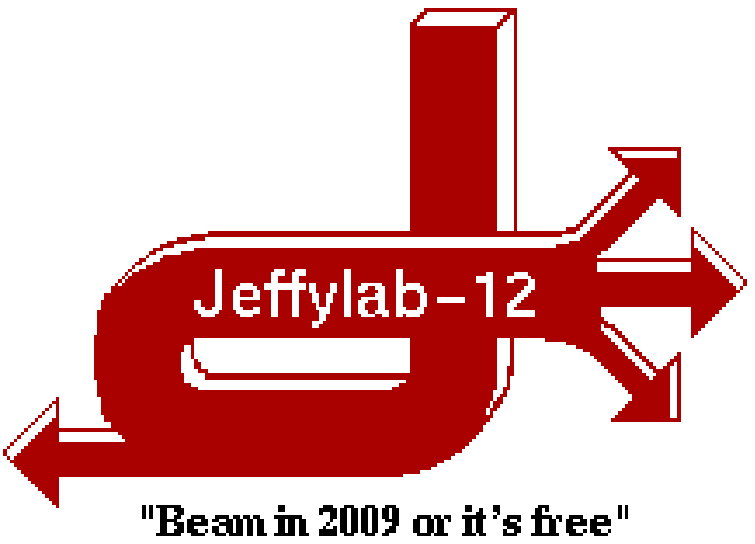}
}
\end{figure}
This work is supported in part by the U.~S. Department of Energy, Office of
Nuclear Physics, under contract W-31-109-ENG-38.  It is also supported by a
wonderful cast and crew working at or with Jefferson Lab, my home away from
home.  I have to sincerely and gratefully acknowledge all those who have
worked so hard to make Jefferson lab what it is today, and to prepare for
an equally important and exciting future.

\bibliography{HUGSsubmission}

\end{document}